\documentclass[aip,cha,reprint,
secnumarabic,
nofootinbib, tightenlines,
nobibnotes, showkeys, showpacs,
groupedaddress
]{revtex4-1}

                        \newif\ifpaper \newif\ifPDF               
                        \newif\ifOUP \newif\ifdraft               
                        \newif\ifdasbuch                          
                        \newif\ifsolutions \newif\ifblog          
                        \blogtrue                                 
                        \draftfalse      
                        \dasbuchtrue 
                        \solutionstrue 
                        \paperfalse\PDFtrue 
                        \OUPfalse                

\usepackage{float}
\usepackage{natbib}
\usepackage{amsmath,amsfonts,amssymb,amsbsy,amscd,amsgen}
\usepackage{dcolumn}
\usepackage{bm}
\usepackage[pdftex]{graphicx}
\usepackage{url}
\usepackage{subfigure}
\usepackage{ifthen}

\usepackage[pdftex,colorlinks]{hyperref} 

        \newcommand{\wwwcb}[1]{       
                  {\tt \href{http://ChaosBook.org#1}
              {ChaosBook.org#1}}}

       \newcommand{\arXiv}[1]{
              {\tt \href{http://arXiv.org/abs/#1}{\goodbreak arXiv:#1}}}

\newcommand{\beq}{\begin{equation}}
\newcommand{\continue}{\nonumber \\ }
\newcommand{\nnu}{\nonumber}
\newcommand{\eeq}{\end{equation}}
\newcommand{\ee}[1] {\label{#1} \end{equation}}
\newcommand{\bea}{\begin{eqnarray}}

\newcommand{\eea}{\end{eqnarray}}
\newcommand{\barr}{\begin{array}}
\newcommand{\earr}{\end{array}}

\newcommand{\rf}     [1] {~\cite{#1}}
\newcommand{\refref} [1] {ref.~\cite{#1}}

\newcommand{\refrefs}[1] {refs.~\cite{#1}}

\newcommand{\refeq}  [1] {(\ref{#1})}

\newcommand{\reffig} [1] {figure~\ref{#1}}
\newcommand{\reffigs} [2] {figures~\ref{#1} and~\ref{#2}}
\newcommand{\refFig} [1] {Figure~\ref{#1}}

\newcommand{\refsect}[1] {sect.~\ref{#1}}

\newcommand{\ie}{{i.e.}}        

\newcommand{\statesp}{state space}

\newcommand{\braket}[2]
		   {\langle{#1}\vphantom{#2}|\vphantom{#1}{#2}\rangle}

\newcommand{\BER}[1]{{\mbox{\footnotesize BER}}} 

\newcommand{\pS}{\ensuremath{{\cal M}}}          
\newcommand{\ssp}{\ensuremath{x}}                
\newcommand{\PoincS}{\ensuremath{{\cal P}}}  

\newcommand\map{f}                  
\newcommand\flow[2]{{f^{#1}(#2)}}
\newcommand{\vel}{\ensuremath{v}}   

\newcommand{\cLe}{complex Lorenz equations}
\newcommand{\cLf}{complex Lorenz flow}

\newcommand{\CLf}{Complex Lorenz flow}
\newcommand{\RerCLor}{\rho_1}    
\newcommand{\ImrCLor}{\rho_2}    

\newcommand\period[1]{{\ensuremath{T_{#1}}}}         

\newcommand{\cycle}[1]{\ensuremath{\overline{#1}}}

\newcommand{\po}{periodic orbit}

\newcommand{\rpo}{rela\-ti\-ve periodic orbit}
\newcommand{\eqv}{equi\-lib\-rium}

\newcommand{\eqva}{equi\-lib\-ria}

\newcommand{\reqv}{rela\-ti\-ve equi\-lib\-rium}
\newcommand{\reqva}{rela\-ti\-ve equi\-lib\-ria}

\newcommand{\PoincSec}{Poincar\'e section}
\newcommand{\reducedsp}{reduced state space}

\newcommand{\slice}{slice}
\newcommand{\Slice}{Slice}
\newcommand{\mslices}{method of slices}

\newcommand{\chartBord}{chart border}
\newcommand{\ChartBord}{Chart border}
\newcommand{\poincBord}{section border}

\newcommand{\template}{template} 

\newcommand{\pSRed}{\ensuremath{\hat{\cal M}}} 
\newcommand{\sspRed}{\ensuremath{\hat{\ssp}}}    
\newcommand{\velRed}{\ensuremath{\hat{\vel}}}    

\newcommand{\slicep}{{\ensuremath{\sspRed'}}}   
\newcommand{\sliceTan}[1]{\ensuremath{t'_{#1}}}    
\newcommand{\groupTan}{\ensuremath{t}}    
\newcommand{\Group}{\ensuremath{G}}         
\newcommand{\Lg}{\ensuremath{\mathbf{T}}}   
\newcommand{\LieEl}{\ensuremath{g}}  

\newcommand{\zeit}{\ensuremath{t}}  
\newcommand{\sspRSing}{\ensuremath{\sspRed^\ast}} 	

\newcommand{\gSpace}{\ensuremath{{\bf \phi}}}   
\newcommand{\velRel}{\ensuremath{c}}    
\newcommand{\phaseVel}{phase velocity}      

\newcommand{\On}[1]{\ensuremath{\textrm{O}(#1)}}
\newcommand{\SOn}[1]{\ensuremath{\textrm{SO}(#1)}}         

\newcommand{\shift}{\ensuremath{d}}

\newcommand{\dmn}{-dimensional}  

\newcommand{\REQV}[2]{\ensuremath{TW_{#1#2}}} 

\renewcommand{\reffig} [1] {Fig.~\ref{#1}}
\renewcommand{\reffigs} [2] {Figs.~\ref{#1} and~\ref{#2}}
\renewcommand{\refFig} [1] {Fig.~\ref{#1}}

\newcommand{\Template}{Template}
\newcommand\mapRed{\ensuremath{\hat{f}}}     

\renewcommand{\zeit}{\ensuremath{\tau}}  

\ifpaper 
       \renewcommand{\arXiv}[1]{ {\tt arXiv:#1}}
       
\else 
       
       \renewcommand{\arXiv}[1]{
              {\tt \href{http://arXiv.org/abs/#1}{arXiv:#1}}}
\fi 

\newcommand{\wurst}{wurst}
\newcommand{\Wurst}{Wurst}

\newcommand{\NSe}{Navier-Stokes equations}
\newcommand{\Reynolds}{\textit{Re}}  
\newcommand{\cohStr}{coherent structure}

\renewcommand{\shift}{\ensuremath{\ell}}

\newcommand{\Norm}[1]{\|{#1}\|}

\begin{document}

\title[High-dimensional cartography]
{Cartography of high-dimensional flows: A visual guide to sections and slices}

\author{Predrag Cvitanovi{\'c}}
\email{predrag@gatech.edu.}
\author{Daniel Borrero-Echeverry}
\author{Keith M. Carroll}
\author{Bryce Robbins}
\author{Evangelos Siminos}
\affiliation{
 Center for Nonlinear Science and School of Physics,
 Georgia Inst. of Technology,
 Atlanta, GA  30332, USA
}

\date{\today}
\date{July 21, 2012}

    \begin{abstract}
Symmetry reduction by the method of slices quotients the continuous
symmetries of chaotic flows by replacing the original state space by a
set of charts, each covering a neighborhood of a dynamically important
class of solutions, qualitatively captured by a `template'. Together
these charts provide an atlas of the symmetry-reduced `slice' of state
space, charting the regions of the manifold explored by the trajectories
of interest. Within the slice, relative equilibria reduce to equilibria
and relative periodic orbits reduce to periodic orbits. Visualizations of
these solutions and their unstable manifolds reveal their interrelations
and the role they play in organizing turbulence/chaos.
    \end{abstract}

\pacs{02.20.-a, 05.45.-a, 05.45.Jn, 47.27.ed, 47.52.+j, 83.60.Wc}

\keywords{
symmetry reduction,
equivariant dynamics,
relative equilibria,
relative periodic orbits,
slices,
moving frames
}
\maketitle

    \begin{quotation}
Today, it is possible to take a stroll through the high-dimensional
\statesp\ of hydrodynamic turbulence and observe that turbulent
trajectories are guided by close passes to invariant solutions of the
\NSe. Charting how close these passes are is a geometer's task, but in
order to place them on a map, one first has to deal with families of
solutions equivalent under the symmetries of a given flow.
Evolution in time decomposes the \statesp\ into a `spaghetti' of time
trajectories. Continuous spatial symmetries foliate it like the layers
of an onion. In this visual tour of dynamics, we use a low-dimensional
flow to illustrate how this tangle can be unraveled (symmetry reduction),
and how to pick a single representative point for each trajectory
(section it) and  group orbit (slice it). Once the symmetry induced
degeneracies are out of the way, one can identify and describe the
prominent turbulent structures by a taxonomy of invariant building
blocks (numerically exact solutions of the \NSe, finite sets of \reqva\
and infinite hierarchies of \rpo s) and describe the dynamics in terms of
near passes to their heteroclinic connections.
    \end{quotation}

\section{Introduction}
\label{s:intro}

Over the last decade, new insights into the dynamics of moderate
\Reynolds\ turbulent flows%
\rf{FE03,WK04,science04,Kerswell05}
have been gained through visualizations of
their $\infty$\dmn\ \statesp s by means of dynamically invariant,
representation independent coordinate frames constructed from physically
prominent unstable {\cohStr s},\rf{GHCW07} hereafter referred to as {\em
\template s}. Navigating and
charting the geometry of these extremely high\dmn\ \statesp s
necessitates a reexamination of two of the basic tools of the theory of
dynamical systems: \PoincSec s and symmetry reduction.

In quantum-mechanical calculations, one always starts out by making sure
that the Hamiltonian has been brought to its
symmetry-reduced block-diagonal, irreducible form;
anything else would be sheer masochism.
As the dynamical theory of turbulence is still in its infancy,
symmetry reduction is not yet a common practice in processing  turbulence
data collected in experimental measurements and numerical simulations.
{
Symmetry reduction of nonlinear flows is much trickier than the more
familiar theory of irreducible representations for linear problems such
as quantum mechanics, so most of our sketches illustrate the simplest
case, the 1-parameter compact continuous group \SOn{2} symmetry.}

We show here how to bring the numerical or experimental data to a
symmetry-reduced format \emph{before} any further analysis of it takes
place.
Our tool of choice is the linear implementation of the
\mslices.\rf{rowley_reconstruction_2000,BeTh04,SiCvi10,FrCv11}
Here, we extend this local method to a global reduction of a turbulent
flow by defining local `charts', their borders, and the ridges that glue
these linear tiles into an atlas that spans the ergodic \statesp\ region
of interest.
{
While `charts' and `atlases' are standard tools in geometry, the
prescription for explicit construction of a symmetry-reduced \statesp\
presented here is, to our knowledge, new.}
We explain the key geometrical ideas in simple but
illustrative settings, eschewing the fluid dynamical and group
theoretical technicalities.

\begin{figure}
   \centering
  \setlength{\unitlength}{0.20\textwidth}
(a)
  \begin{picture}(1,0.98239821)%
    \put(0,0){\includegraphics[width=\unitlength]{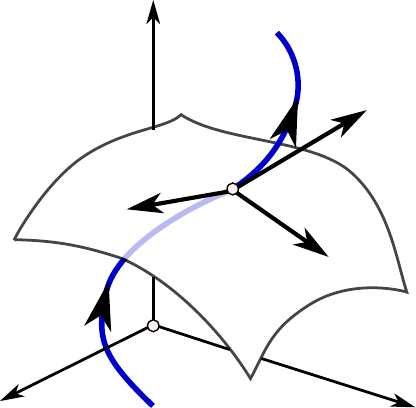}}%
    \put(0.91612064,0.70682767){\color[rgb]{0,0,0}\makebox(0,0)[lb]{\smash{$\vel$}}}%
    \put(0.48698745,0.90266503){\color[rgb]{0,0,0}\makebox(0,0)[lb]{\smash{$\ssp(\zeit)$}}}%
    \put(0.2624318,0.5347756){\color[rgb]{0,0,0}\makebox(0,0)[lb]{\smash{$\groupTan_1$}}}%
    \put(0.80471037,0.38188675){\color[rgb]{0,0,0}\makebox(0,0)[lb]{\smash{$\groupTan_2$}}}%
    \put(0.538343,0.25344355){\color[rgb]{0,0,0}\makebox(0,0)[lb]{\smash{$\pS_\ssp$}}}%
    \put(0.47864531,0.56060893){\color[rgb]{0,0,0}\makebox(0,0)[lb]{\smash{$\ssp$}}}%
  \end{picture}%
~~(b)
  \begin{picture}(1,0.9071451)%
    \put(0,0){\includegraphics[width=\unitlength]{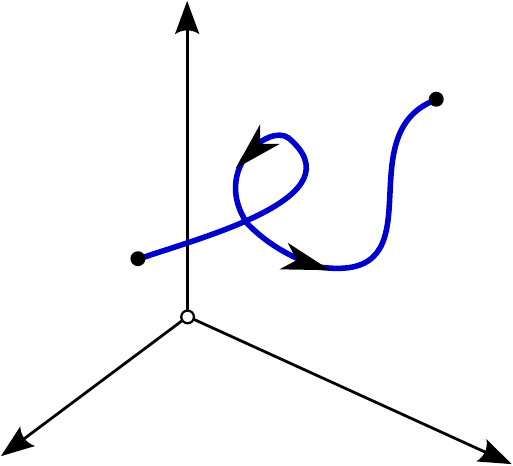}}%
    \put(0.81413243,0.74633551){\color[rgb]{0,0,0}\makebox(0,0)[lb]{\smash{$\ssp(\zeit)$}}}%
    \put(0.14085867,0.30233715){\color[rgb]{0,0,0}\makebox(0,0)[lb]{\smash{$\ssp(0)$}}}%
  \end{picture}%
\\
(c)
	\begin{picture}(1,0.88265338)%
    \put(0,0){\includegraphics[width=\unitlength]{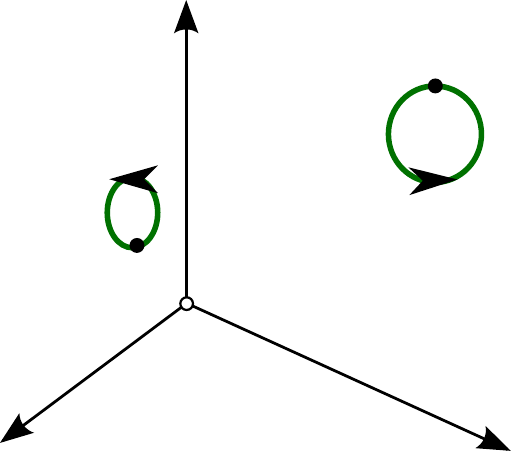}}%
    \put(0.17017404,0.30123416){\color[rgb]{0,0,0}\makebox(0,0)[lb]{\smash{$\ssp(0)$}}}%
    \put(0.10473094,0.59736144){\color[rgb]{0,0,0}\makebox(0,0)[lb]{\smash{$\pS_{\ssp(0)}$}}}%
    \put(0.80073137,0.42842331){\color[rgb]{0,0,0}\makebox(0,0)[lb]{\smash{$\pS_{\ssp(\zeit)}$}}}%
    \put(0.81421682,0.75061198){\color[rgb]{0,0,0}\makebox(0,0)[lb]{\smash{$\ssp(\zeit)$}}}%
	\end{picture}%
~~(d)
	\begin{picture}(1,0.90708568)%
    \put(0,0){\includegraphics[width=\unitlength]{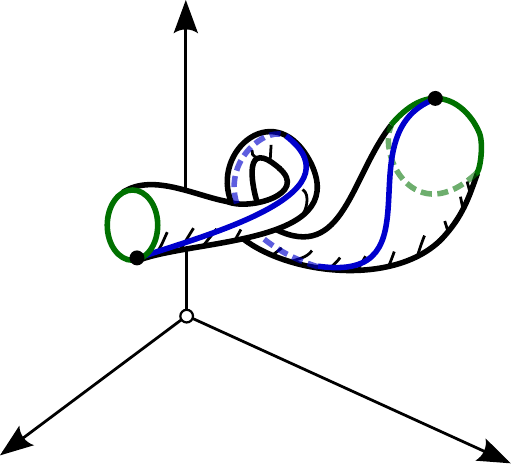}}%
    \put(0.81871555,0.75240026){\color[rgb]{0,0,0}\makebox(0,0)[lb]{\smash{$\ssp(\zeit)$}}}%
    \put(0.12337724,0.29209515){\color[rgb]{0,0,0}\makebox(0,0)[lb]{\smash{$\ssp(0)$}}}%
	\end{picture}%
\caption{\label{fig:A27wurst}
   (a)
In the presence of an $N$-continuous parameter symmetry, each \statesp\
point $\ssp$ owns $(N\!+\!1)$ tangent vectors: one $\vel(\ssp)$ along the
time flow $\ssp(\zeit)$, and the $N$ group tangents  $\groupTan_1(\ssp),
\, \groupTan_2(\ssp) ,\,\cdots, \groupTan_N(\ssp)$ along infinitesimal
symmetry shifts, tangent to the {$N$\dmn} group orbit $\pS_\ssp$.
    (b)
{Each point has a unique trajectory (blue) under time evolution.}
    (c)
{Each point also belongs to a group orbit (green) of
symmetry-related points. For \SOn{2}, this is topologically a circle.
Any two points on a group orbit are physically equivalent, but may
lie far from each other in \statesp.}
    (d)
{Together, time-evolution and group actions trace out a \wurst\ of
physically equivalent solutions.}
}
\end{figure}

Let us begin by defining a {dynamical system} comprised of a flow
$\map^t$ and the \statesp\ $\pS$ on which it acts. If a group $\Group$ of
continuous transformations acts on a continuous time flow, each \statesp\
point owns a set of tangent vectors (\reffig{fig:A27wurst}\,(a)).
Integrated in time, the velocity vector $\vel(\ssp)$ traces out a {\em
trajectory} $\flow{\zeit}{\ssp}$ (\reffig{fig:A27wurst}\,(b)). Applying
the continuous transformations traces out a {group orbit}
\(
\pS_\ssp = \{\LieEl\,\ssp \mid \LieEl \in {\Group}\}
\,
\) 
(\reffig{fig:A27wurst}\,(c)). Together, time evolution and group actions
trace out a complicated smooth manifold, hereafter affectionately
referred to as a {\em \wurst} (see Figs.~\ref{fig:A27wurst}\,(d),
\ref{fig:CLf01group}\,(b) and \ref{fig:sliceimage}), which we shall here
teach you how to slice.

A flow is said to have symmetry $\Group$ if the form of evolution
equations $\dot{\ssp} = \vel(\ssp)$ is left invariant,
\(
\vel(\ssp)=\LieEl^{-1} \, \vel(\LieEl \, \ssp)
\,,
\) 
by the set of transformations $\LieEl \in {\Group}$.
{
If a flow has symmetry, the simplest solutions are highly symmetric
invariant \eqva\ and \reqva\ studied in bifurcation-theory approaches to
the onset of turbulence. Physicists love symmetries,\rf{Kerswell12} but
nature often prefers solutions of no symmetry: while the flow equations
may be invariant under $\Group$, turbulent solutions are not. The highly
symmetric solutions often lie far from the regions of \statesp\ explored
by turbulence\rf{ACHKW11} and thus are of limited usefulness in
understanding its dynamics. In contrast, the \rpo s studied here are
embedded in the turbulence, and capture its geometry and statistics.}

We can make headway in unraveling the tangle
of 1\dmn\ time trajectories with the notion of recurrence. To quantify how
close the state of the system at a given time is to a previously
visited state, we need the notion of distance between two points in
\statesp. The simplest (but far from the only, or the most natural) is
the Euclidean norm
\beq
  \Norm{\ssp-\ssp'}^2  = \braket{\ssp-\ssp'}{\ssp-\ssp'} =
\sum_{i=1}^d
(\ssp-\ssp')_i^2
\,.
\ee{innerproduct}
For experimental data, a better norm,
for example, might be a distance between digitized
images. While in this paper we simply assume that a norm is given, its
importance cannot be overstated: the construction of invariant, PDE
discretization independent \statesp\ coordinates,\rf{GHCW07} the symmetry
reduction by minimization of the distance between group orbits undertaken
in what follows, and the utility of the charts so constructed all depend on a
well-chosen notion of distance in the high\dmn\ \statesp s we are
charting here.

Given a notion of distance, we can talk about a `neighborhood', an open
set of nearby states with qualitatively similar dynamics. Our main task
in what follows will be to make this precise by defining a chart over a
neighborhood and its borders. Given distances and neighborhoods, the next
key notion is  \emph{measure}, or how likely a typical trajectory is to
visit a given neighborhood. After some observations of a given turbulent
flow, one can identify a set of \emph{\template
s},\rf{rowley_reconstruction_2000} {points} $\slicep{}^{(j)}$,
$j=1,2,\cdots$ in the \statesp\ representative of the most frequently
revisited features of the flow.

Our goals here are two-fold:
(i) In \refsect{s:cut}, we review the method of \PoincSec s, with
    emphasis on two particular aspects that are applicable to
    high\dmn\ flows: the construction of multiple local linear
    charts and the determination of their borders.
(ii) In \refsect{s:symm}, we discuss the effect of continuous symmetries on
    nonlinear flows, and in \refsect{s:slice} we use the lessons learned
    from our discussion of \PoincSec s to aid us in the reduction of
    continuous symmetries, and, thus, enable us to commence a systematic
    charting of the long-time dynamics of high\dmn\ flows
    (\refsect{s:chart}).

\section{Section}
\label{s:cut}

In the {\em \PoincSec} method, one records the coordinates $\sspRed_n$ of
the trajectory $\ssp(\zeit)$ at the instants $\zeit_n$ when it traverses
a fixed oriented hypersurface $\PoincS$ of codimension 1. For the
high\dmn\ flows that we have in mind, the practical choice is a
hyperplane, the only type of \PoincSec\ (from now on, just a
\emph{section}) that we shall consider here. One can choose a section
such that it contains a \template\ of interest. Properly oriented, such a
section can capture important features of the flow in the neighborhood of
the section-fixing \template.

\begin{figure}
 \begin{center}
 \setlength{\unitlength}{0.20\textwidth}
(a)
  \begin{picture}(1,0.8736435)%
    \put(0,0){\includegraphics[width=\unitlength]{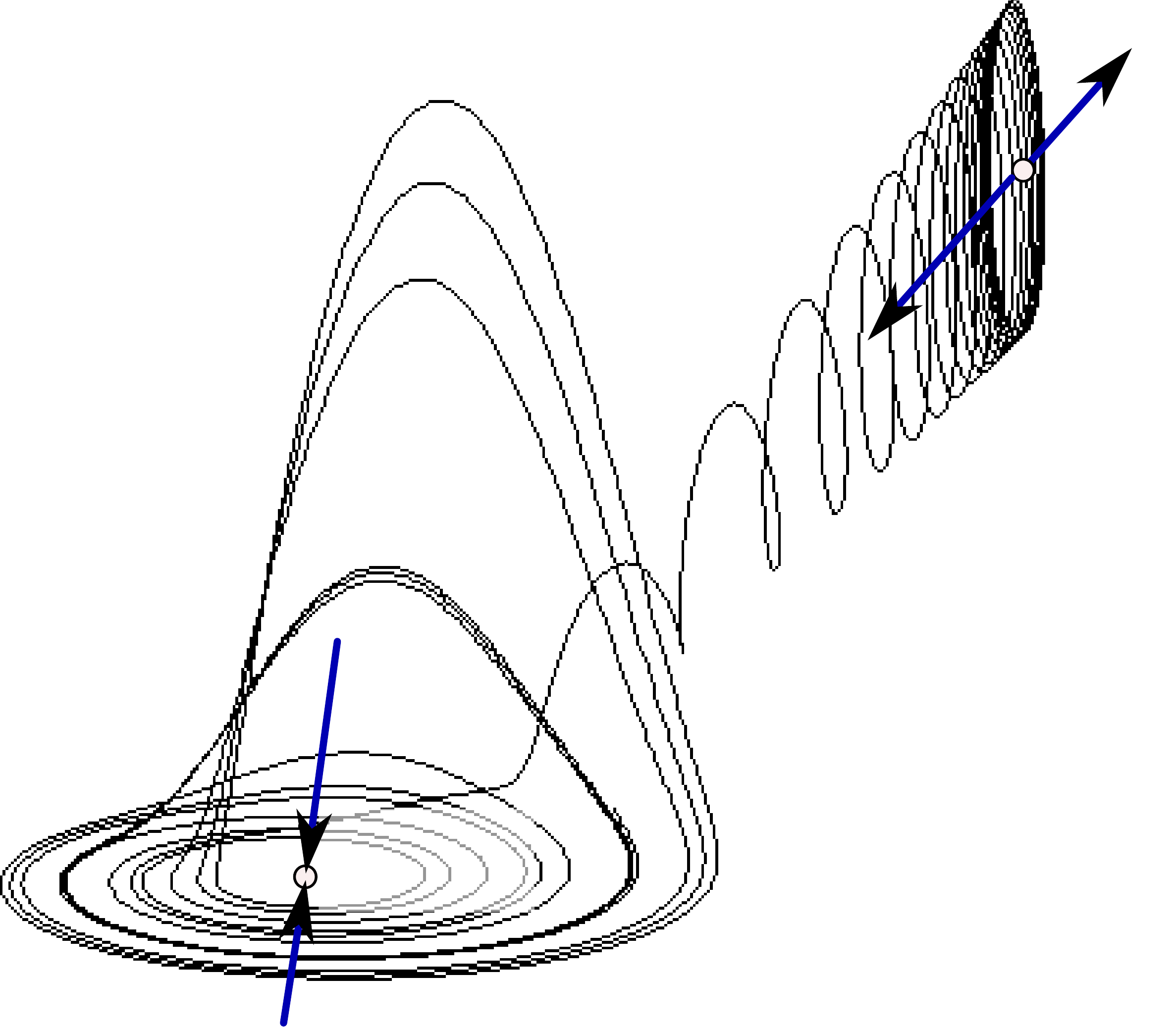}}%
    \put(0.27213233,0.11036704){\color[rgb]{0,0,0}\makebox(0,0)[lb]{\smash{$\slicep{}^{(-)}$}}}%
    \put(0.90759454,0.68935432){\color[rgb]{0,0,0}\makebox(0,0)[lb]{\smash{$\slicep{}^{(+)}$}}}%
  \end{picture}%
(b) 
  \begin{picture}(1,0.82646416)%
    \put(0,0){\includegraphics[width=\unitlength]{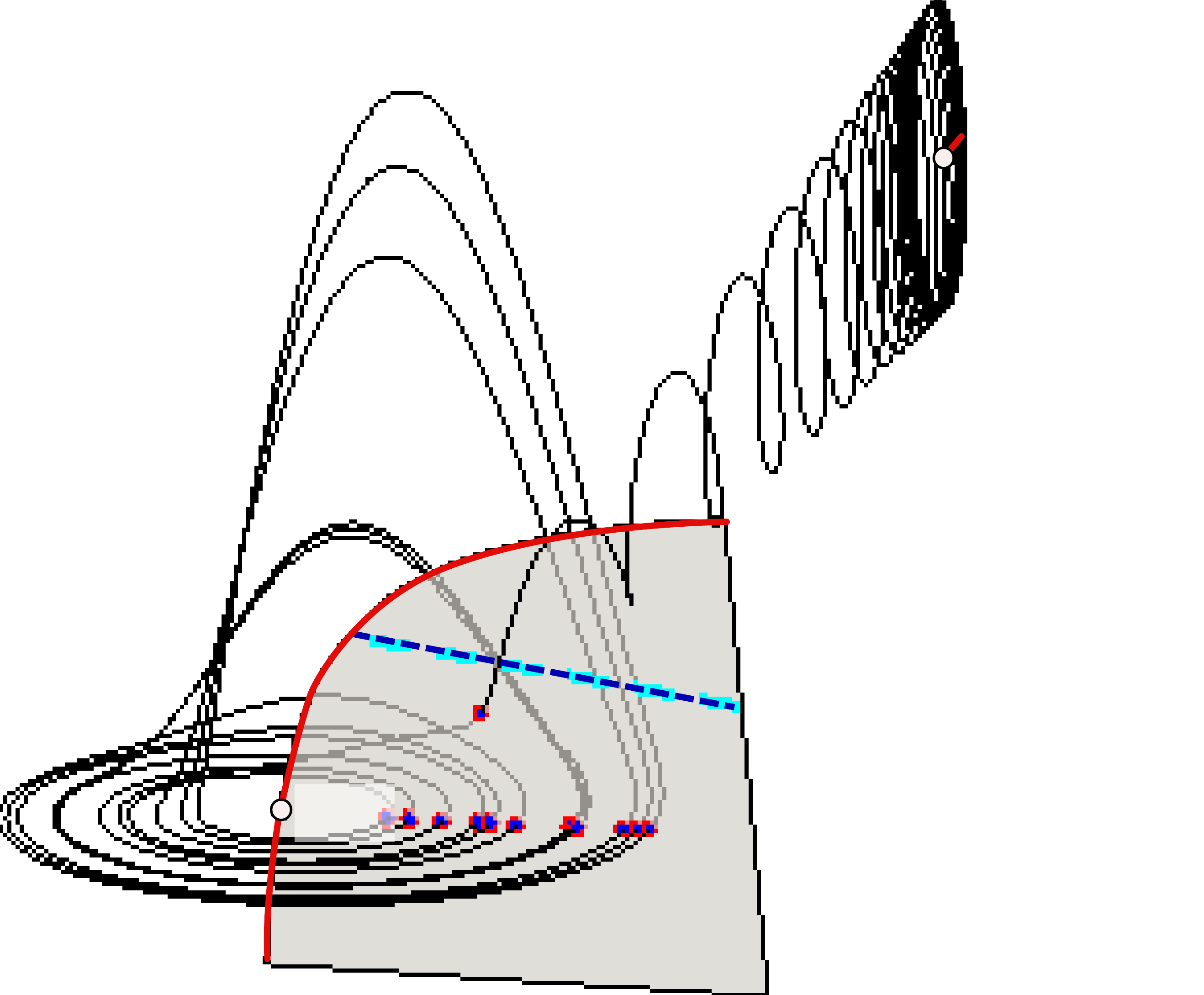}}%
    \put(0.2468894,0.13781942){\color[rgb]{0,0,0}\makebox(0,0)[lb]{\smash{$\slicep{}^{(-)}$}}}%
  \end{picture}%
\\
(c)  
  \begin{picture}(1,0.82646416)%
    \put(0,0){\includegraphics[width=\unitlength]{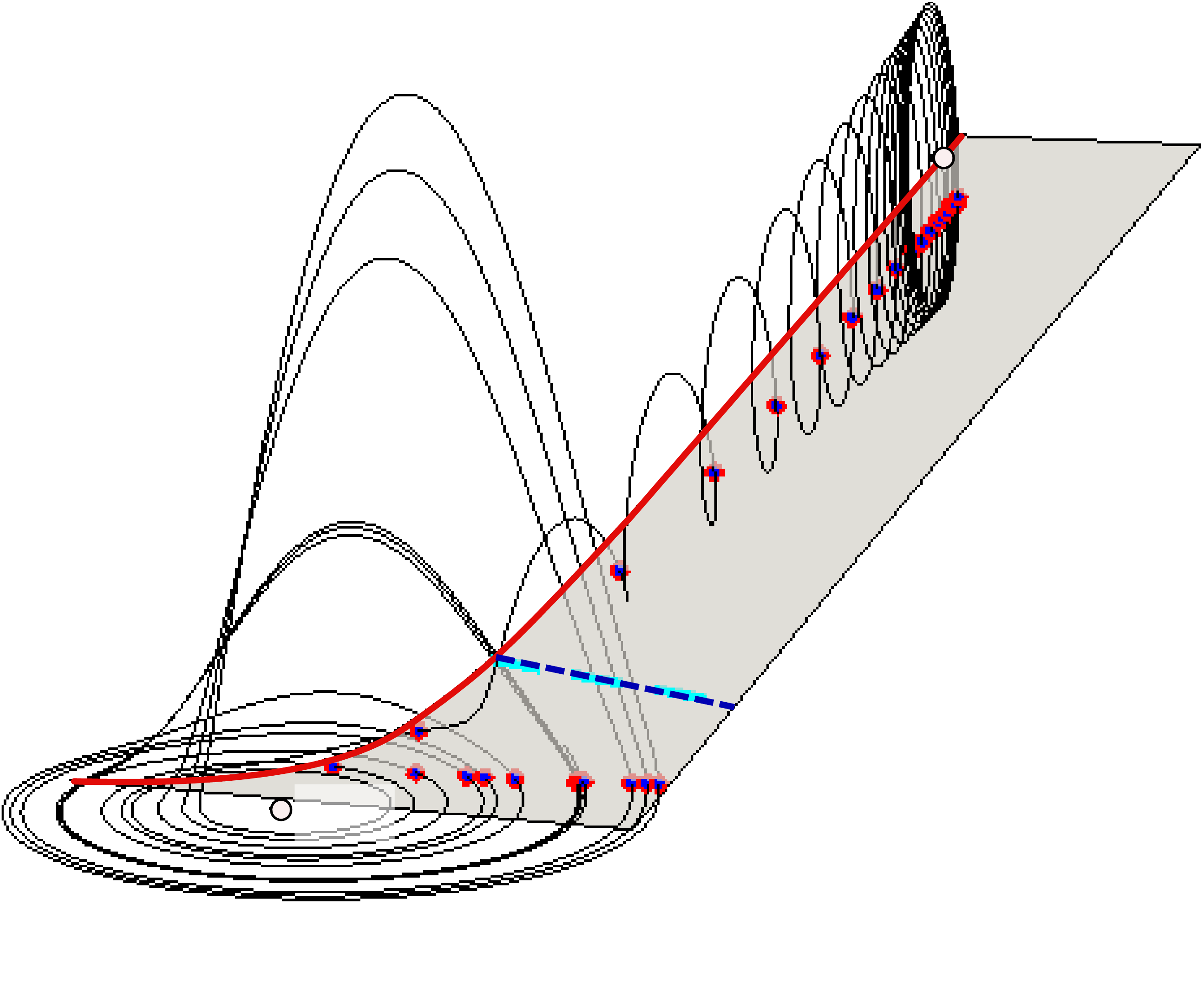}}%
    \put(0.81012394,0.72899125){\color[rgb]{0,0,0}\makebox(0,0)[lb]{\smash{$\slicep{}^{(+)}$}}}%
  \end{picture}%
(d)  
  \begin{picture}(1,0.82646416)%
    \put(0,0){\includegraphics[width=\unitlength]{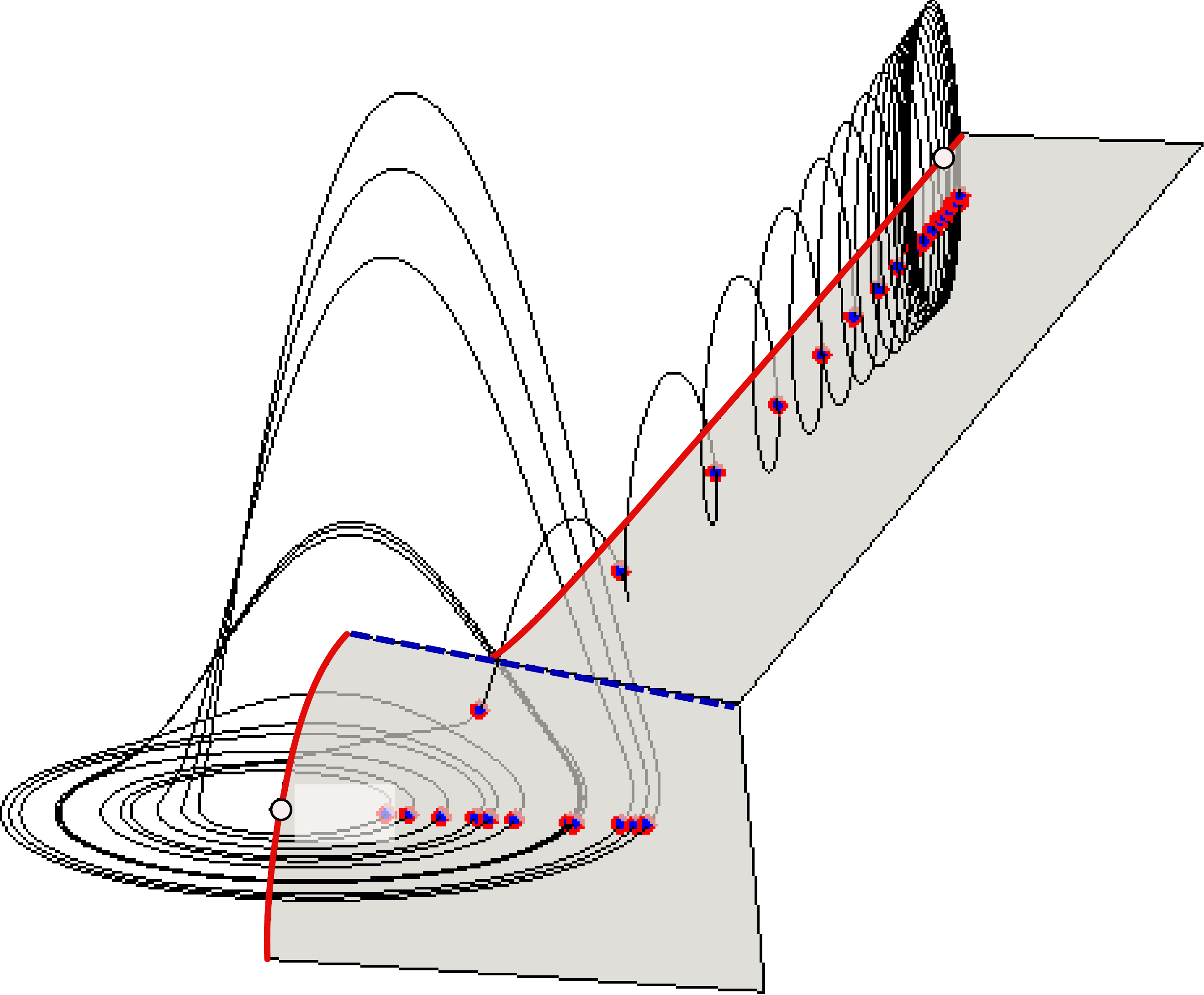}}%
    \put(0.2468894,0.13781942){\color[rgb]{0,0,0}\makebox(0,0)[lb]{\smash{$\slicep{}^{(-)}$}}}%
    \put(0.81012394,0.72884196){\color[rgb]{0,0,0}\makebox(0,0)[lb]{\smash{$\slicep{}^{(+)}$}}}%
  \end{picture}%
 \end{center}
    \caption{
2-chart atlas for R\"ossler flow.
(a)
  The inner {\eqv} $\slicep{}^{(-)}$  is a (spiral-out) saddle-focus with
  a 2\dmn\ unstable manifold and a 1\dmn\ stable manifold.
  The outer {\eqv} $\slicep{}^{(+)}$ is a (spiral-in) saddle-focus, with
  a 2\dmn\ stable manifold (basin boundary for initial conditions
  that either fall into the  chaotic attractor, or escape to infinity)
  and a 1\dmn\ unstable manifold.
(b)
 Chart $\PoincS_{-}$ of the $\slicep{}^{(-)}$ neighborhood carved out of a
 \PoincSec\ plane through the inner {\eqv} $\slicep{}^{(-)}$ and its
 stable eigenvector, with \poincBord\ drawn as the solid red line.
 Note the ridge
 (dashed blue line): the chart stops at the ridge.
(c)
  Chart $\PoincS_{+}$ (here viewed from below) is bounded by \poincBord\
  (solid red line) of a section through the outer {\eqv}
  $\slicep{}^{(+)}$  and its unstable eigenvector. The chart stops at the
  ridge (dashed blue line), and it does not intersect the strange
  attractor.
(d)
  A two-chart atlas of R\"ossler flow, with charts $\PoincS_{-}$ and
  $\PoincS_{+}$ oriented and combined so that the ridge (intersection of
  the two sections, indicated by the dashed blue line in the three
  figures) lies approximately between the \template s. Section
  hyperplanes beyond this ridge do not belong to the atlas.
    }
\label{fig:RoessTrjs}
\end{figure}

But how far does this neighborhood extend? The answer is that the section
captures neighboring trajectories as long as it cuts them transversally;
it fails the moment the velocity field at a point $\sspRSing$ fails to
pierce the section. At these locations, the velocity either vanishes
(\eqv) or is orthogonal to the section normal $\hat{n}$,
\beq
    \hat{n} \cdot \vel(\sspRSing) = 0
\,,\qquad
    \sspRSing \in \cal{S}
\,.
\ee{eq:sspRSing}
For a smooth flow in $d$ dimensions such points form a smooth $(d\!-\!2)$\dmn\
\emph{\poincBord} ${\cal S} \subset \PoincS$, which encloses the open
neighborhood of the {\template} characterized by qualitatively similar
flow. We shall refer to this region of the section as a \emph{chart} of the
{\template} neighborhood (see \reffig{fig:RoessTrjs}). Beyond the border,
the flow pierces the section in the `wrong' direction and the dynamics
are qualitatively different.

As an example consider the R\"ossler system\rf{ross},
\index{R\"ossler system}
\beq
\begin{split}
  \dot{x} &= -y \,-\,z \\
  \dot{y} &= x + a y \\
  \dot{z} &= b + z (x - c)
  \,,
  \label{eq:Rossler}
\end{split}
\eeq
where $a = b = 0.2$ and $c = 5.7$. This flow has two prominent invariant
states, the `inner' and the `outer' unstable \eqva\ $\slicep{}^{(-)}$ and
$\slicep{}^{(+)}$ (see \refFig{fig:RoessTrjs}\,(a)) , which we choose as
{\em \template s} for our sections.

We orient the sections so the plane $\PoincS_{-}$ contains $\slicep{}^{(-)}$ and its 1\dmn\
stable eigenvector (\reffig{fig:RoessTrjs}\,(b)),
and the other section $\PoincS_{+}$ contains $\slicep{}^{(+)}$ and its 1\dmn\ unstable
eigenvector (\reffig{fig:RoessTrjs}\,(c)), thus
capturing the local spiral-in, spiral-out dynamics. The remaining freedom
to rotate each section can be used to orient them in such a way that the
ridge (the intersection of the two sections) lies approximately between
the two templates (\reffig{fig:RoessTrjs}\,(d)). Choosing sections is a
dark art: in the example at hand the dynamics of interest is captured by
the two charts - if that were not the case, one would have had to
interpolate, by inserting a third chart between them.

%

For R\"ossler flow, the border condition \refeq{eq:sspRSing} yields a
quadratic condition in 3 dimensions, so the \poincBord s\ drawn in
\reffig{fig:RoessTrjs}\,(b) and \reffig{fig:RoessTrjs}\,(c) are conic
sections. The two charts meet at a ridge, and together do a pretty good
job as the 2-chart atlas of the interesting R\"ossler dynamics. Due to the
extreme contraction rate of the attractor, its intersection with
the section in
\reffig{fig:RoessTrjs}\,(b) is for all practical purposes 1\dmn, and the
associated return map yields all \po s of the 3\dmn\ flow.\rf{DasBuch}

In 3 dimensions everything ---sections, ridges, \poincBord s--- can be
drawn and the chart fits on a 2\dmn\ sheet of papyrus.
{
But what about for hydrodynamic flows where the
dimensionality $d$ of the \statesp\ is very large? The point of the
cartographical enterprize undertaken here is that while it is impossible
to visualize the $(d\!-\!2)$\dmn\ {\poincBord} of the $(d\!-\!1)$\dmn\
slab that is now our chart,\rf{GibsonMovies} a point is a point and a
line is a line in a projection from any number of dimensions, so a
trajectory crossing of either a section or a {\poincBord} can be easily
determined and visualized in any dimension.}

To summarize:
Evolution in time decomposes the \statesp\ into a spaghetti of 1\dmn\
trajectories $\ssp(\zeit)$, each fixed by picking a single point $\ssp(0)$
on it. A well chosen set of {section charts} of codimension 1 allows us to
`quotient' the continuous time parameter $\zeit$, and reveal the
dynamically important transverse structure of the flow's stable/unstable
manifolds. For unstable trajectories one needs, in addition, a notion of
recurrence to the section. The set of points $\{\sspRed_n\} =
\{\ssp(\zeit_n)\}$,  separated by short time flights in between sections,
captures the transverse dynamics without losing any information about the
chaotic flow. We can thus chart interesting regions of \statesp\ by
picking a sufficient number of \template s and using them to construct
charts of their neighborhoods, each bounded by \poincBord s and ridges.

{
We close this section with a remark on what sections \emph{are not}:}
A \PoincSec\ is {not} a projection onto a lower\dmn\ space (in
sense that a photograph is a 2\dmn\ projection of a 3\dmn\
space). Rather, it is a local change of coordinates to a direction along
the flow $\vel(\sspRed)$, and the remaining coordinates transverse to it.
No information about the flow is lost; the full space trajectory
$\ssp(\zeit)$ can always be reconstructed by integration from its point
$\sspRed$ in the section.

\section{Dancers and drifters}
\label{s:symm}

 \begin{figure}
 \begin{center}
  \setlength{\unitlength}{0.20\textwidth}
(a)
  \begin{picture}(1,0.52454249)%
    \put(0,0){\includegraphics[width=\unitlength]{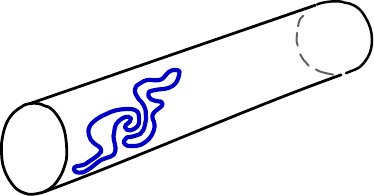}}%
    \put(0.61583231,0.13683004){\color[rgb]{0,0,0}\makebox(0,0)[lb]{\smash{$z$}}}%
    \put(0.00611823,0.27217453){\color[rgb]{0,0,0}\makebox(0,0)[lb]{\smash{$\theta$}}}%
  \end{picture}%
(b)
  \begin{picture}(1,0.52454249)%
    \put(0,0){\includegraphics[width=\unitlength]{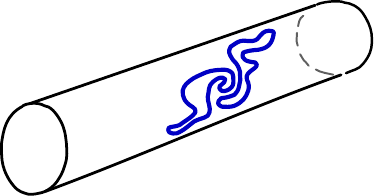}}%
    \put(0.61583231,0.13683004){\color[rgb]{0,0,0}\makebox(0,0)[lb]{\smash{$z$}}}%
    \put(0.00611823,0.27217453){\color[rgb]{0,0,0}\makebox(0,0)[lb]{\smash{$\theta$}}}%
  \end{picture}%
\\
(c)
  \begin{picture}(1,0.52454249)%
    \put(0,0){\includegraphics[width=\unitlength]{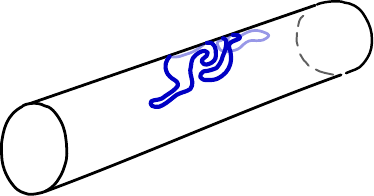}}%
    \put(0.61583231,0.13683004){\color[rgb]{0,0,0}\makebox(0,0)[lb]{\smash{$z$}}}%
    \put(0.00611823,0.27217453){\color[rgb]{0,0,0}\makebox(0,0)[lb]{\smash{$\theta$}}}%
  \end{picture}%
(d)
  \begin{picture}(1,0.52454249)%
    \put(0,0){\includegraphics[width=\unitlength]{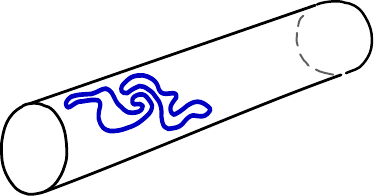}}%
    \put(0.61583231,0.13683004){\color[rgb]{0,0,0}\makebox(0,0)[lb]{\smash{$z$}}}%
    \put(0.00611823,0.27217453){\color[rgb]{0,0,0}\makebox(0,0)[lb]{\smash{$\theta$}}}%
  \end{picture}%
 \end{center}
 \caption[$\On{2}_\theta \times \SOn{2}_z$ symmetry of flow in a stream-wise
          periodic pipe]{
{
A symmetry relates physically equivalent states; a pipe flow
solution translated or rotated is also a solution.
(a) An instantaneous state of the  fluid is indicated by a `swirl' -
here the reader has to imagine a particular instantaneous velocity field
across the entire pipe. The same state may be rigidly
(b) translated by downstream shift $\shift$
    (fluid states are $\SOn{2}_z$ equivariant in a stream-wise periodic pipe),
(c) translated by $\shift$ and rotated azimuthally by $\gSpace$
    (the two states are $\SOn{2}_\theta \times \SOn{2}_z$ equivariant), and
(d) reflected and rotated azimuthally by $\gSpace$
    (the two states are $\On{2}_\theta$ equivariant).
Some symmetry-related states may also be connected by time evolution.
A \emph{\reqv} is a solution of the equations of motion that retains its
    shape while rotating and traveling downstream with constant
    {\phaseVel} $\velRel$.
A \emph{\rpo} $\pS_p$ is a \emph{time dependent}, shape-changing
    state of the fluid that after a period $\period{p}$ reemerges as (b), (c), or (d),
    the initial state translated by $\shift_p$, rotated by $\gSpace_p$
    and possibly also azimuthally reflected.}
 }\label{fig:A27-pipeSymms}
 \end{figure}
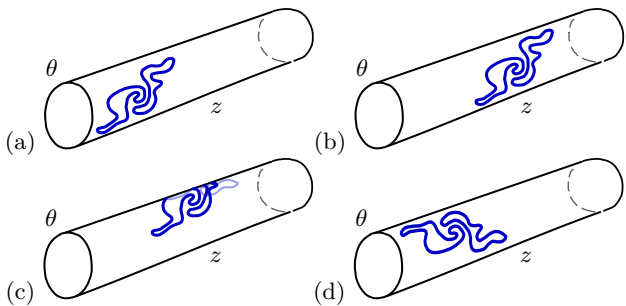

What is a symmetry? A visualization of the fluid dynamics of a pipe flow,
\reffig{fig:A27-pipeSymms}, affords an intuitive illustration. Solutions
of pipe flow remain physically the same under azimuthal rotations and
stream-wise translations (which become \SOn{2} rotations in numerical
stream-wise periodic pipes), but rotated and shifted solutions may
correspond to distant points in \statesp.

Each \SOn{2} group orbit is topologically a circle, but it traces out a
complicated \statesp\ curve composed of many Fourier modes that are
nonlinearly coupled and thus of comparable magnitude. Together, the two
\SOn{2} symmetries of numerical pipe flow sweep out contorted and hard to
visualize $T^2$ tori (see \refref{ACHKW11}), so we shall illustrate the
key ideas by a much simpler example, the $\SOn{2}$-equivariant Gibbon and
McGuinness\rf{GibMcCLE82,FowlerCLE82} \cLe\ of geophysics and laser
physics,
\bea
	\dot{x}_1 &=& -\sigma x_1 + \sigma y_1
        \,,\qquad
	\dot{x}_2 \,=\, -\sigma x_2 + \sigma y_2
        \continue
	\dot{y}_1 &=& (\RerCLor-z) x_1 - \ImrCLor x_2 -y_1-e y_2 \continue
	\dot{y}_2 &=& \ImrCLor x_1 + (\RerCLor-z) x_2 + e y_1- y_2\continue
	\dot{z} \; &=& -b z + x_1 y_1 + x_2 y_2
    \,.
\label{eq:CLeR}
\eea
{
Here all coordinates and parameters are real. In our calculations}
the parameters are set to those used by Siminos'\rf{SiminosThesis}
$\RerCLor=28,\, \ImrCLor=0,\, b=8/3,\, \sigma=10,\, e= 1/10$.
{
The \cLe\ are an example of a simple dynamical system with a continuous
(but no discrete) symmetry, equivariant under \SOn{2} rotations by
\bea
\LieEl(\gSpace)
    &=&
\exp{({\gSpace} \Lg)}
	 \,=\,
  \left(\barr{ccccc}
  \cos \gSpace  & \sin \gSpace  & 0 & 0 & 0 \\
 -\sin \gSpace  & \cos \gSpace  & 0 & 0 & 0 \\
 0 & 0 &  \cos \gSpace & \sin \gSpace   & 0 \\
 0 & 0 & -\sin \gSpace & \cos \gSpace   & 0 \\
 0 & 0 & 0             & 0              & 1
    \earr\right)
\nnu 
\eea
The group is 1\dmn\ and compact, its elements parameterized by $\gSpace
\mbox{ mod } 2\pi$.}
For historical background, Poincar\'e return maps, symbolic dynamics and
in-depth investigation of the model, see \refrefs{SiminosThesis,SiCvi10}.

\begin{figure}
  	\begin{center}
  	\setlength{\unitlength}{0.20\textwidth}
  (a)
  	\begin{picture}(1,1.07802818)%
    	\put(0,0){\includegraphics[width=\unitlength]{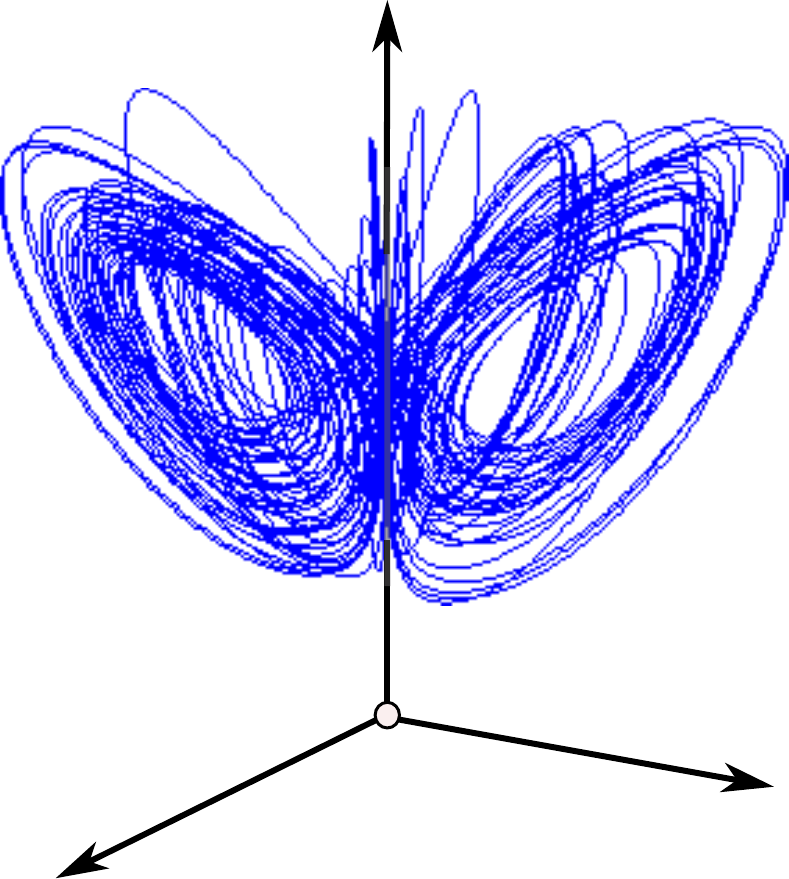}}%
    	\put(0.55152995,1.0139628){\color[rgb]{0,0,0}\makebox(0,0)[lb]{\smash{$z$}}}%
    	\put(0.05573445,0.0739776){\color[rgb]{0,0,0}\makebox(0,0)[lb]{\smash{$x_1$}}}%
    	\put(0.90013492,0.16491708){\color[rgb]{0,0,0}\makebox(0,0)[lb]{\smash{$x_2$}}}%
  	\end{picture}%
  (b)
  	\begin{picture}(1,1.06440474)%
    	\put(0,0){\includegraphics[width=\unitlength]{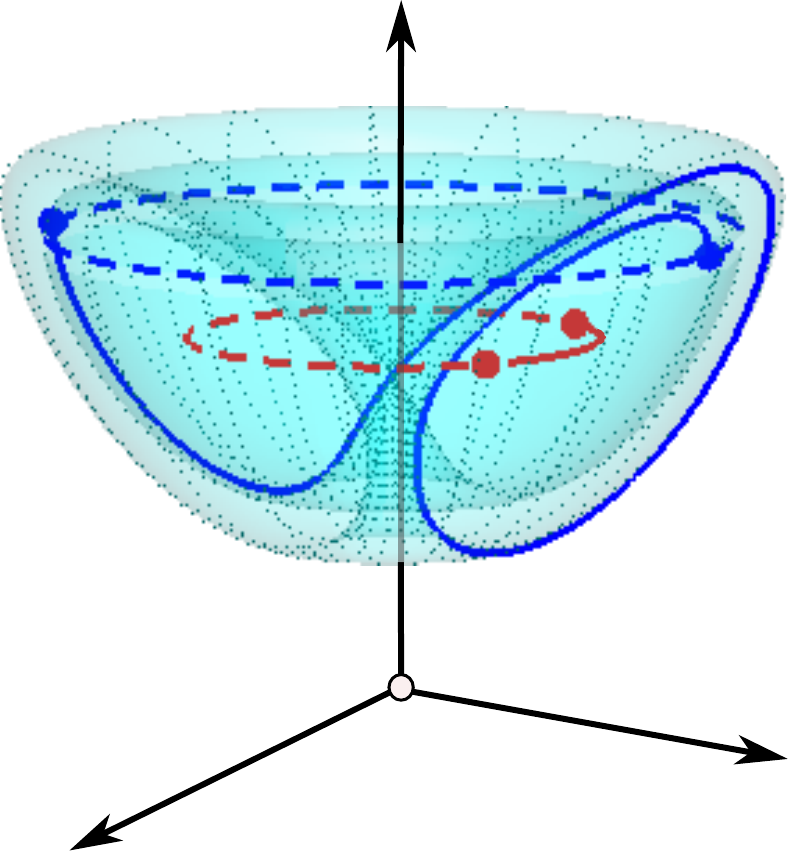}}%
   		\put(0.55961552,1.00214901){\color[rgb]{0,0,0}\makebox(0,0)[lb]{\smash{$z$}}}%
   		\put(0.07008555,0.07304272){\color[rgb]{0,0,0}\makebox(0,0)[lb]{\smash{$x_1$}}}%
    	\put(0.90381504,0.16283301){\color[rgb]{0,0,0}\makebox(0,0)[lb]{\smash{$x_2$}}}%
  	\end{picture}
\\
(c)   \begin{picture}(1,0.94310243)%
    \put(0,0){\includegraphics[width=\unitlength]{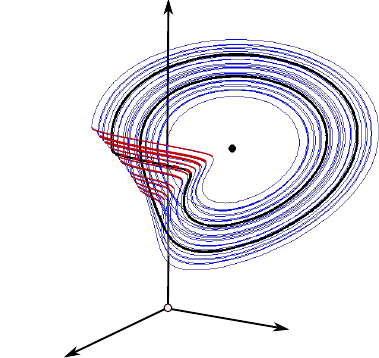}}%
    \put(0.48564392,0.89244183){\color[rgb]{0,0,0}\makebox(0,0)[lb]{\smash{$z$}}}%
    \put(0.07181137,0.03185892){\color[rgb]{0,0,0}\makebox(0,0)[lb]{\smash{$y_2$}}}%
    \put(0.77031544,0.100183){\color[rgb]{0,0,0}\makebox(0,0)[lb]{\smash{$x_2$}}}%
  \end{picture}%
(d)   \begin{picture}(1,1.05662086)%
    \put(0,0){\includegraphics[width=\unitlength]{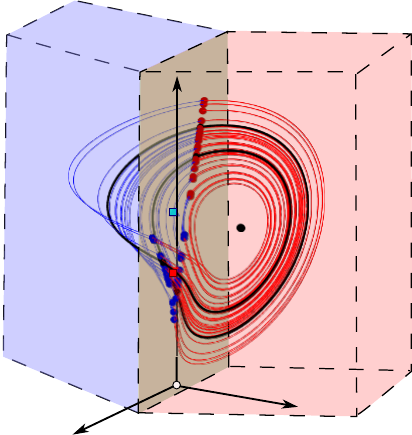}}%
    \put(0.47706962,0.83002768){\color[rgb]{0,0,0}\makebox(0,0)[lb]{\smash{$z$}}}%
    \put(0.08719004,0.02997825){\color[rgb]{0,0,0}\makebox(0,0)[lb]{\smash{$y_2$}}}%
    \put(0.73025395,0.09287946){\color[rgb]{0,0,0}\makebox(0,0)[lb]{\smash{$x_2$}}}%
  \end{picture}
    \end{center}
  \caption
  [\CLf: $\cycle{01}$ {\rpo} group orbit]{
  \CLf, $d=5 \to 3$~dimensional $\{x_1,x_2,z\}$ projections:
  (a)
  The strange attractor.
  (b)
  The initial \reqv\ $\REQV{}{1}$ point is shown by the red dot, and its
  group orbit / trajectory by the dashed red line. One period of the
  $\cycle{01}$ {\rpo} is shown by the solid blue line. The group orbit of
  its (arbitrary) starting point is shown by the dashed blue line: after
  one period the trajectory has returned to the group orbit but with a
  different phase. The \wurst, \ie, the group orbit of the $\cycle{01}$
  trajectory (dark blue) is shown by the cyan surface. Following
  $\cycle{01}$ for 15 more periods (faint dotted lines) starts filling
  out this torus; in that time, the slowly drifting \reqv\ $\REQV{}{1}$
  has advanced to the next red dot (red line).
Symmetry-reduced \cLf, $d=4 \to 3$~dimensional $\{x_2,y_2,z\}$ projections:
 (c)
 Strange attractor from frame (a) reduced to a single \slice\ hyperplane,
 using $\REQV{}{1}$ as the \template. $\cycle{01}$ is now a {\po}, shown
 by the solid black line. The dynamics exhibits singular jumps
 (shown in red) due to forbidden crossings of the \chartBord. In contrast
 to the 1\dmn\ \poincBord s of \reffig{fig:RoessTrjs}, here the \chartBord s
 are 3\dmn\ and hard to visualize.
 (d)
The 2-chart atlas (see the sketch of \reffig{fig:A29-1ridge}) of the same
strange attractor encounters no \chartBord s and exhibits no
singularities. The trajectory changes colors from red to blue as it
crosses between the \slice\ hyperplanes of $\slicep{}^{(1)}$ and
$\slicep{}^{(2)}$. The ridge (shown in brown) acts as a \PoincSec\
$\PoincS$ with red or blue ridge points $\sspRed^*$ marking the direction
of the crossing. The charts are 4\dmn, the ridge 3\dmn, so the colored
blocks and planes are only cartoon drawings of their projections onto the
2\dmn\ figure.
  }
\label{fig:CLf01group}
\end{figure}

The strange attractor of the \cLf\ , in its
present state, is a complete mess (\reffig{fig:CLf01group}\,(a)). Solutions tend to drift along
continuous symmetry directions, with the physically important
shape-changing dynamics hidden from view.

The ultimate drifter, the signature invariant solution that signals
the presence of a continuous symmetry is a {\em \reqv} (traveling wave,
rotational wave, etc.), a trajectory whose velocity field lies within the
group tangent space,
\(
\vel(\ssp) = c \cdot \groupTan(\ssp)
\,,
\) 
and whose time evolution is thus confined to the group orbit (see
\reffig{fig:CLf01group}\,(b)); think of an unchanging body carried by a
stream.

A {\em \rpo} behaves more like a dancer. $\pS_p$ is a trajectory that
recurs exactly
\beq
\ssp(\zeit) = \LieEl_p \, \ssp(\zeit + \period{p} )
    \,,\qquad
\ssp(\zeit) \in \pS_p
    \,,
\ee{RPOrelper1}
after a fixed {relative period} $\period{p}$, but shifted by a fixed
group action ${\LieEl_p}$ that maps the endpoint $\ssp (\period{p}) $ back
into the initial point cycle point $\ssp (0) $; think of a dancer moving
across the stage through a set of motions and then striking her initial
pose,\rf{ShWi06} or study the pipe flow sketches in \reffig{fig:A27-pipeSymms}.

Because the $\SOn{2}$ transformations act on the \cLf\ only through the
simplest $m=1$ Fourier mode, here all group orbits are circles and
appear elliptical in $d=5 \to 3$~dimensions projections. Nevertheless,
even the \wurst\ traced out by one of
the simplest \rpo s
$\cycle{01}$\rf{SiCvi10} (shown in \reffig{fig:CLf01group}\,(b)) is not so
easy to get one's head around: you are looking at a 3\dmn\ projection of
a \emph{torus} embedded in 5 dimensions.

\begin{figure}
 \begin{center}
  \setlength{\unitlength}{0.20\textwidth}
(a)~~
  \begin{picture}(1,0.98655417)%
    \put(0,0){\includegraphics[width=\unitlength]{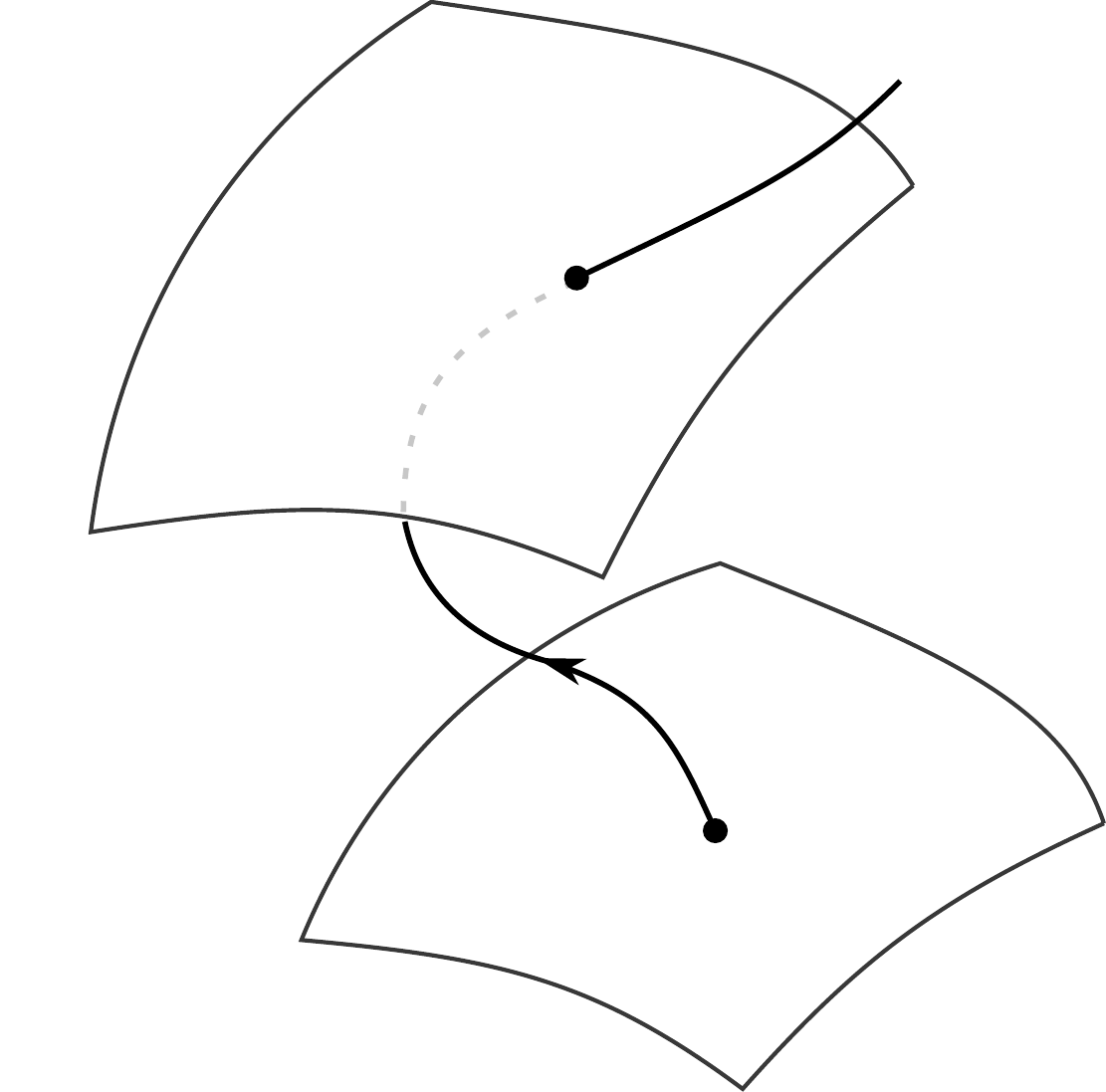}}%
    \put(0.35976094,0.91875614){\color[rgb]{0,0,0}\rotatebox{-16.32889204}{\makebox(0,0)[lb]{\smash{$\pS_{\ssp(\zeit)}$}}}}%
        \put(0.60333631,0.42274457){\color[rgb]{0,0,0}\rotatebox{-30.8073288}{\makebox(0,0)[lb]{\smash{$\pS_{\ssp(0)}$}}}}%
    \put(0.63001383,0.14959019){\color[rgb]{0,0,0}\rotatebox{0.0313674}{\makebox(0,0)[lb]{\smash{$\ssp(0)$}}}}%
    \put(0.4558276,0.64524238){\color[rgb]{0,0,0}\rotatebox{0.0313674}{\makebox(0,0)[lb]{\smash{$\ssp(\zeit)$}}}}%
    \put(0.13110825,0.05766516){\color[rgb]{0,0,0}\rotatebox{0.11031334}{\makebox(0,0)[lb]{\smash{$\pS$}}}}%
  \end{picture}%
~~(b)
  \begin{picture}(1,0.98742208)%
    \put(0,0){\includegraphics[width=\unitlength]{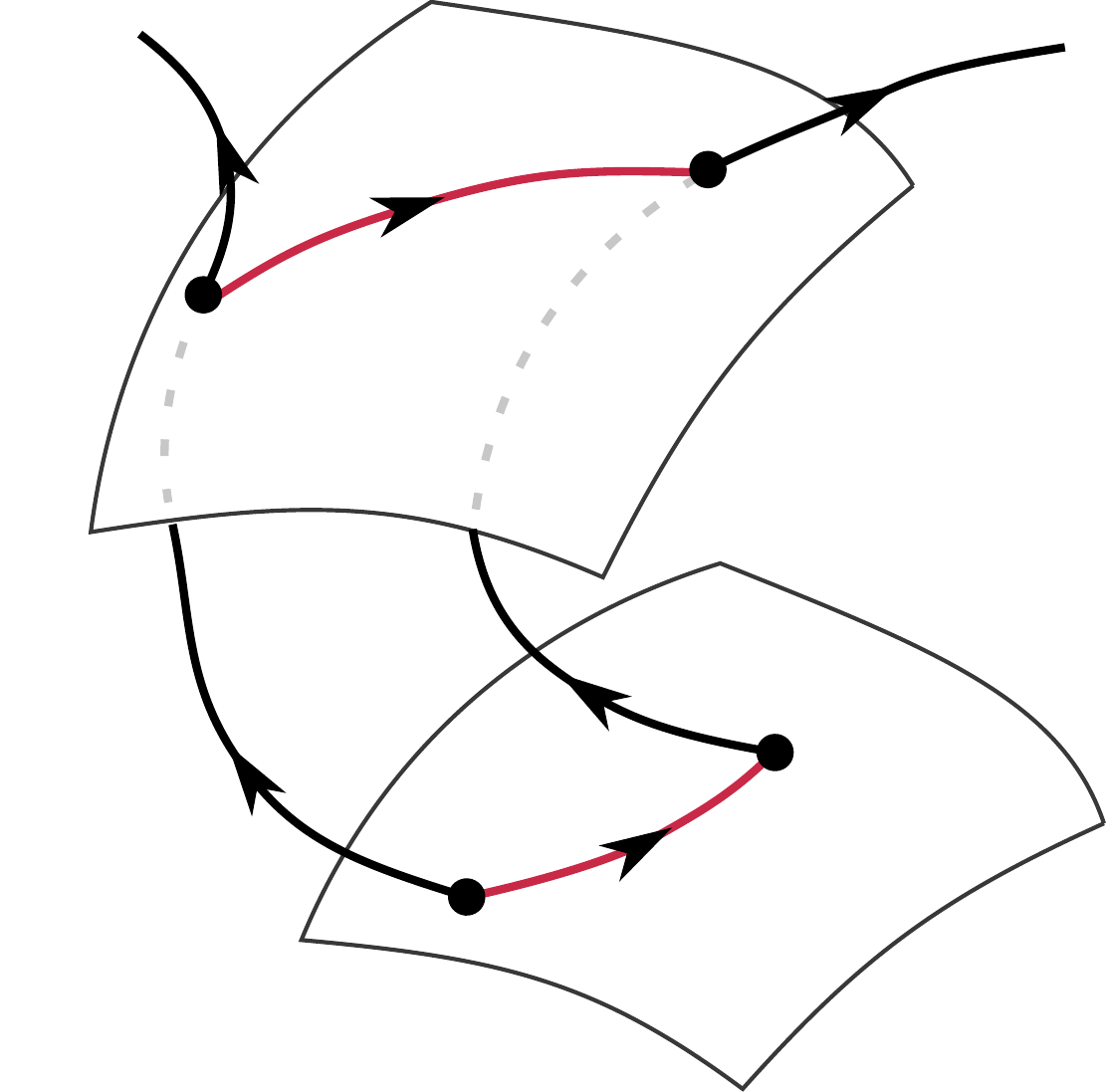}}%
    \put(0.20546387,0.64069775){\color[rgb]{0,0,0}\rotatebox{0.0313674}{\makebox(0,0)[lb]{\smash{$\ssp(\zeit)$}}}}%
    \put(0.7334028,0.27844745){\color[rgb]{0,0,0}\rotatebox{0.0313674}{\makebox(0,0)[lb]{\smash{$\sspRed(0)$}}}}%
    \put(0.52082757,0.7302884){\color[rgb]{0,0,0}\rotatebox{0.0313674}{\makebox(0,0)[lb]{\smash{$\sspRed(\zeit)$}}}}%
    \put(0.35738205,0.88152375){\color[rgb]{0,0,0}\rotatebox{0.0313674}{\makebox(0,0)[lb]{\smash{$\LieEl(\zeit)$}}}}%
    \put(0.48490227,0.11640486){\color[rgb]{0,0,0}\rotatebox{0.0313674}{\makebox(0,0)[lb]{\smash{$\ssp(0)$}}}}%
    \put(0.3947908,0.25892637){\color[rgb]{0,0,0}\rotatebox{0.0313674}{\makebox(0,0)[lb]{\smash{$\LieEl(0)$}}}}%
    \put(0.1047908,0.0792637){\color[rgb]{0,0,0}\rotatebox{0.0313674}{\makebox(0,0)[lb]{\smash{$\pS$}}}}%
  \end{picture}%
\\
(c)
  \begin{picture}(1,1.07315413)%
    \put(0,0){\includegraphics[width=\unitlength]{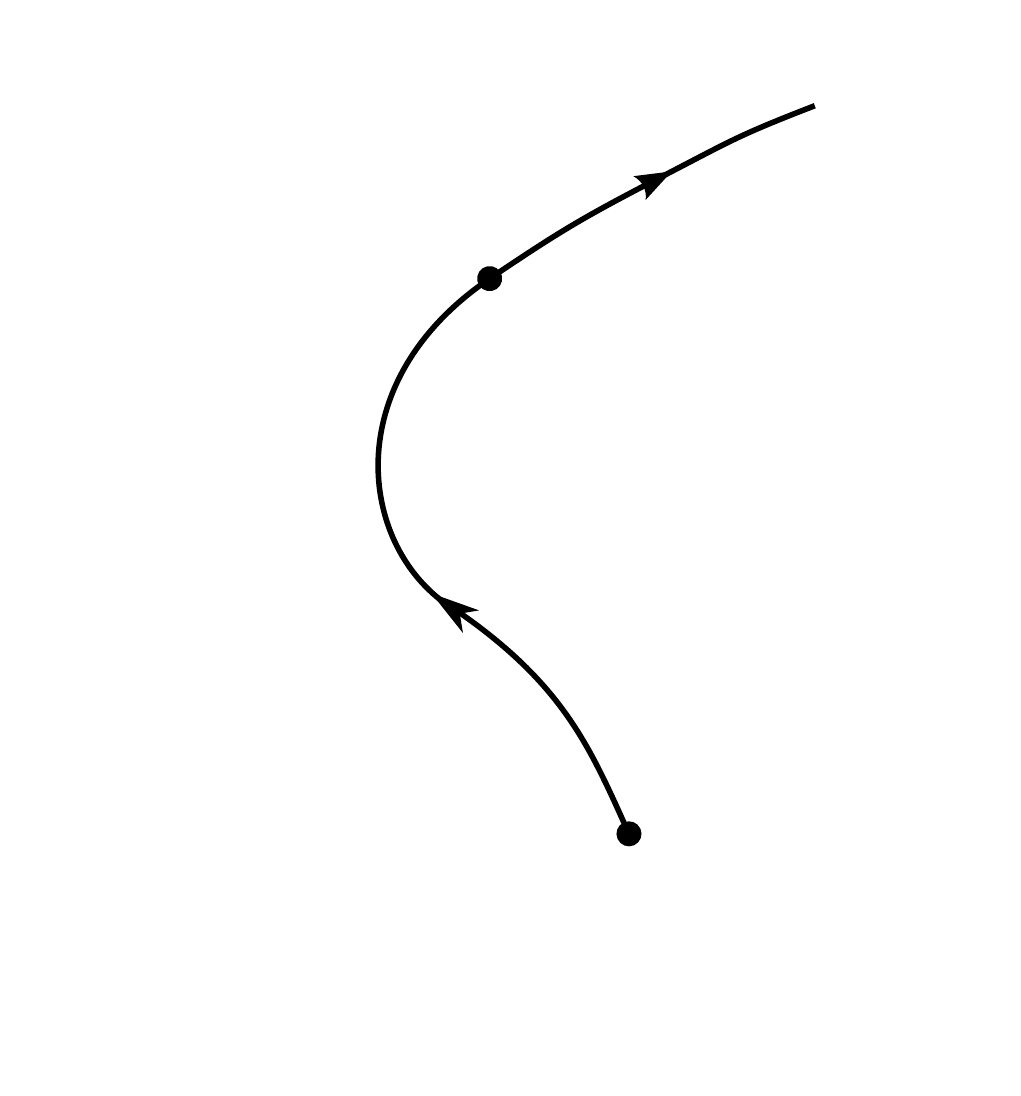}}%
    \put(0.19912369,0.17144733){\color[rgb]{0,0,0}\rotatebox{0.11031334}{\makebox(0,0)[lb]{\smash{$\pSRed$}}}}%
    \put(0.63028127,0.18433598){\color[rgb]{0,0,0}\rotatebox{0.03136739}{\makebox(0,0)[lb]{\smash{$\sspRed(0)$}}}}%
    \put(0.48253394,0.69182305){\color[rgb]{0,0,0}\rotatebox{0.03136739}{\makebox(0,0)[lb]{\smash{$\sspRed(\zeit)$}}}}%
  \end{picture}%
(d)~~~~
  	\begin{picture}(1,0.62007592)%
    	\put(0,0){\includegraphics[width=\unitlength]{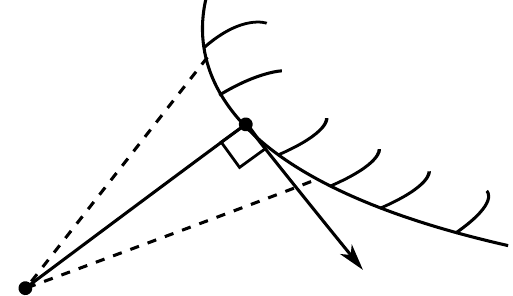}}%
    	\put(0.8274739,0.27313042){\color[rgb]{0,0,0}\makebox(0,0)[lb]{\smash{$\pS_{\slicep}$}}}%
    	\put(0.71822871,0.05712263){\color[rgb]{0,0,0}\makebox(0,0)[lb]{\smash{$\sliceTan{}$}}}%
    	\put(-0.00055527,0.03229441){\color[rgb]{0,0,0}\makebox(0,0)[lb]{\smash{$\sspRed$}}}%
    	\put(0.50966967,0.34820145){\color[rgb]{0,0,0}\makebox(0,0)[lb]{\smash{$\slicep$}}}%
  	\end{picture}
 \end{center}
  \caption{
(a)
The {$N$\dmn} group orbit $\pS_{\ssp(0)}$ of \statesp\ point
$\ssp(0)$ and the group orbit $\pS_{\ssp(\zeit)}$ reached by the
trajectory $\ssp(\zeit)$ a time $\zeit$ later.
(b)
Two physically equivalent trajectories $\ssp(\zeit)$ and $\sspRed(\zeit)$
are related, in general, by an arbitrary, time dependent {\em moving
frame} transformation $\LieEl(\zeit)$, such that
$\ssp(\zeit)=\LieEl(\zeit)\,\sspRed(\zeit)$.
(c)
  {A symmetry reduction scheme $\pS \to \pSRed$ is a rule that
  prescribes $\LieEl(\zeit)$ and thus} replaces a group orbit
  $\pS_{\ssp}\subset\pS$ through \ssp\ by a single point $\sspRed \in
  \pSRed$.
(d)
  {In this paper, $\LieEl(\zeit)$ is fixed variationally by the
  extremal condition \refeq{PCsectQ0} for the point $\sspRed$ on the
  group orbit $\pS_\ssp$ that is nearest to the \template\ $\slicep$.}
  }\label{fig:BeThMovFr}
\end{figure}

To summarize:
continuous symmetries in the dynamics foliate the \statesp\ into an
onion, where each layer is a group orbit (\reffig{fig:BeThMovFr}\,(a)).
How are we to sort out this mess? All the points on a group orbit are
physically equivalent, so we are free to replace a given flow
$\ssp(\zeit)$ by any other \sspRed(\zeit), such that
$\ssp(\zeit)=\LieEl(\zeit)\,\sspRed(\zeit)$ by a {\em moving
frame}\rf{FelsOlver98,FelsOlver99,OlverInv} transformation
$\LieEl(\zeit)$. As long as no symmetry reduction procedure is prescribed,
$\LieEl(\zeit)$ is free: it can be any, in general  time dependent, group
transformation. For example, to film our dancer, we can mount the camera
on a cart moving alongside her. So, in the presence of continuous
symmetries, there are two kinds of motion: those of a dancer,
continuously changing shapes, and those of a drifter, merely shuffling
along the shape invariant directions. We will presently banish the
drifters and just enjoy the dance.

\section{Chart}
\label{s:slice}

Suppose {you are computing a set of} numerically exact invariant solutions
of the \NSe. Do you want to compute the same solution over and
over again, once for every point on the group orbit? No, you would like to
compute it only once. The strategy for picking out that one
representative solution is called \emph{symmetry reduction}. Its goal is
to replace each group orbit by a unique point in a lower\dmn\
symmetry-\reducedsp\ $\pSRed \subset \pS/\Group$, as sketched in
\reffig{fig:BeThMovFr}\,(c).

What is a smart way to go about it? Intuition gained from pipe flow (see
\reffig{fig:A27-pipeSymms}) will again prove helpful. A turbulent flow
exhibits a myriad of unstable structures, each traveling down the pipe
with its own {\phaseVel}. The
\mslices\rf{rowley_reconstruction_2000,BeTh04,SiCvi10,FrCv11} that we now
describe tells you how to pull each solution back into a {\em fixed}
frame called a \emph{\slice} and compare it to your repertoire of
precomputed solutions, or the \template s $\{\slicep{}^{(j)}\}$, using the
poor geometer's version of a geodesic, the principle of the \emph{closest
distance} to each. What follows is similar to the construction of
sections of \refsect{s:cut}; due to the linear action of the symmetry
group, slicing is easier than sectioning, but wholly unfamiliar. This is
why we reviewed the \PoincSec s first. We now offer a pictorial tour of
this (save for one bold incursion\rf{ACHKW11}) hitherto uncharted
territory.

First, pick a \template\ $\slicep$ and use the freedom to shift and
rotate it (\reffig{fig:BeThMovFr}\,(b)) until it overlies, as well as
possible, the state $\ssp$, by minimizing the distance
\beq
\Norm{\ssp - \LieEl(\gSpace)\,\slicep}
\, .
\ee{minDistance}
Now, replace the entire group orbit of $\ssp$ by the closest match to the
template pattern, given by $\sspRed=\LieEl^{-1}\ssp$. From here on, we
will use the hat on $\sspRed$ to indicate the unique point on the group
orbit of $\ssp$ that is closest to the \template\ \slicep. The
symmetry-\reducedsp\ $\pSRed$ is comprised of such closest matches, a
point for each full \statesp\ group orbit.

\begin{figure}
 \begin{center}
  \setlength{\unitlength}{0.30\textwidth}
(a)
  \begin{picture}(1,0.87085079)%
    \put(0,0){\includegraphics[width=\unitlength]{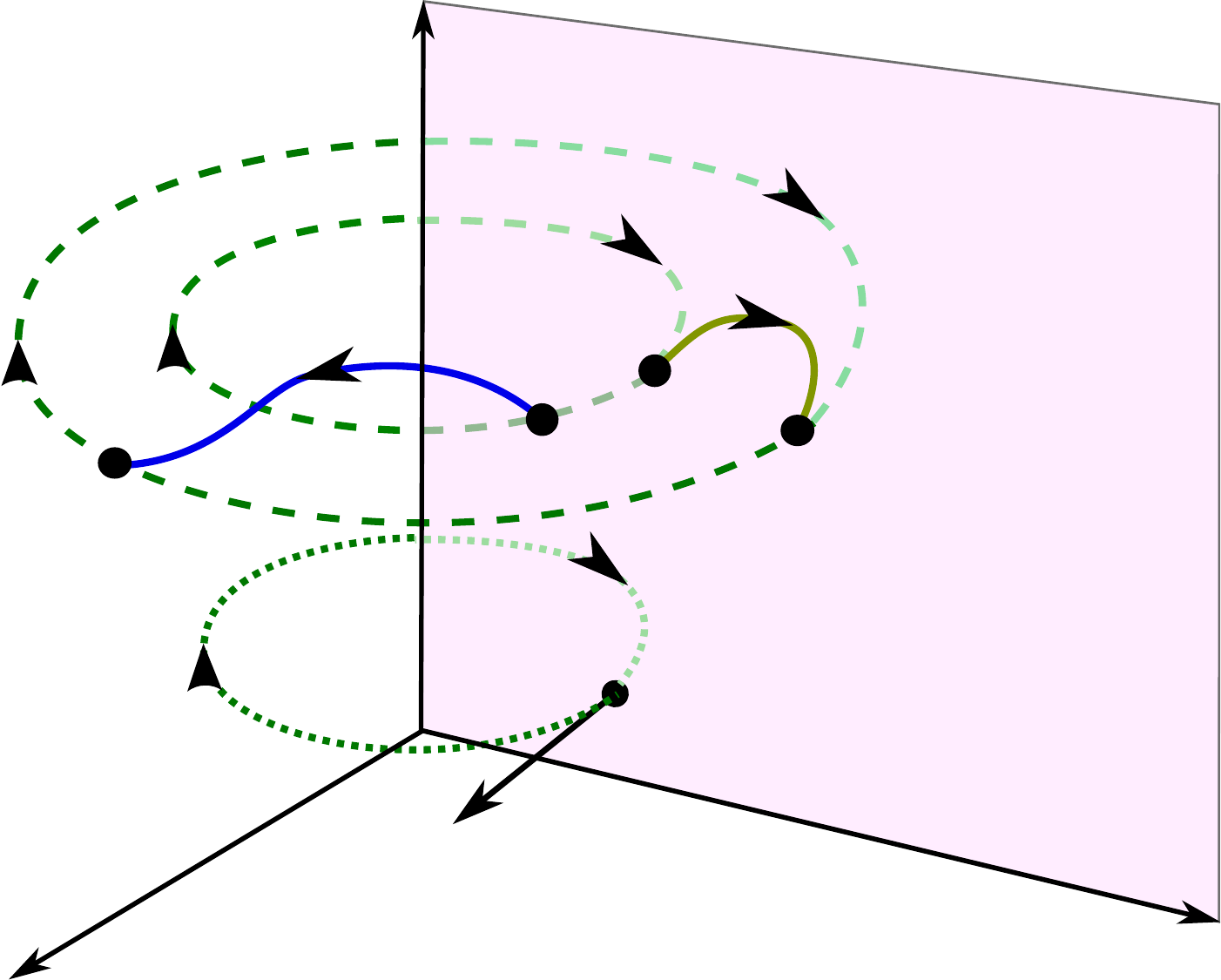}}%
    \put(0.82835155,0.19007659){\color[rgb]{0,0,0}\rotatebox{-14.84025424}{\makebox(0,0)[lb]{\smash{$\pSRed$}}}}%
    \put(0.06577338,0.24688228){\color[rgb]{0,0,0}\rotatebox{0.0313674}{\makebox(0,0)[lb]{\smash{$\LieEl\,\slicep$}}}}%
    \put(0.53023327,0.21593335){\color[rgb]{0,0,0}\rotatebox{0.0313674}{\makebox(0,0)[lb]{\smash{$\slicep$}}}}%
    \put(0.3184954,0.089285){\color[rgb]{0,0,0}\rotatebox{0.0313674}{\makebox(0,0)[lb]{\smash{$\sliceTan{}$}}}}%
    \put(0.00008985,0.35305068){\color[rgb]{0,0,0}\rotatebox{0.0313674}{\makebox(0,0)[lb]{\smash{$\ssp(\zeit)$}}}}%
    \put(0.69766235,0.41412105){\color[rgb]{0,0,0}\rotatebox{0.0313674}{\makebox(0,0)[lb]{\smash{$\sspRed(\zeit)$}}}}%
    \put(0.06716446,0.70280851){\color[rgb]{0,0,0}\rotatebox{0.0313674}{\makebox(0,0)[lb]{\smash{$\LieEl\,\ssp(\zeit)$}}}}%
  \end{picture}%
\\ 
(b)
  \begin{picture}(1,0.87085079)%
    \put(0,0){\includegraphics[width=\unitlength]{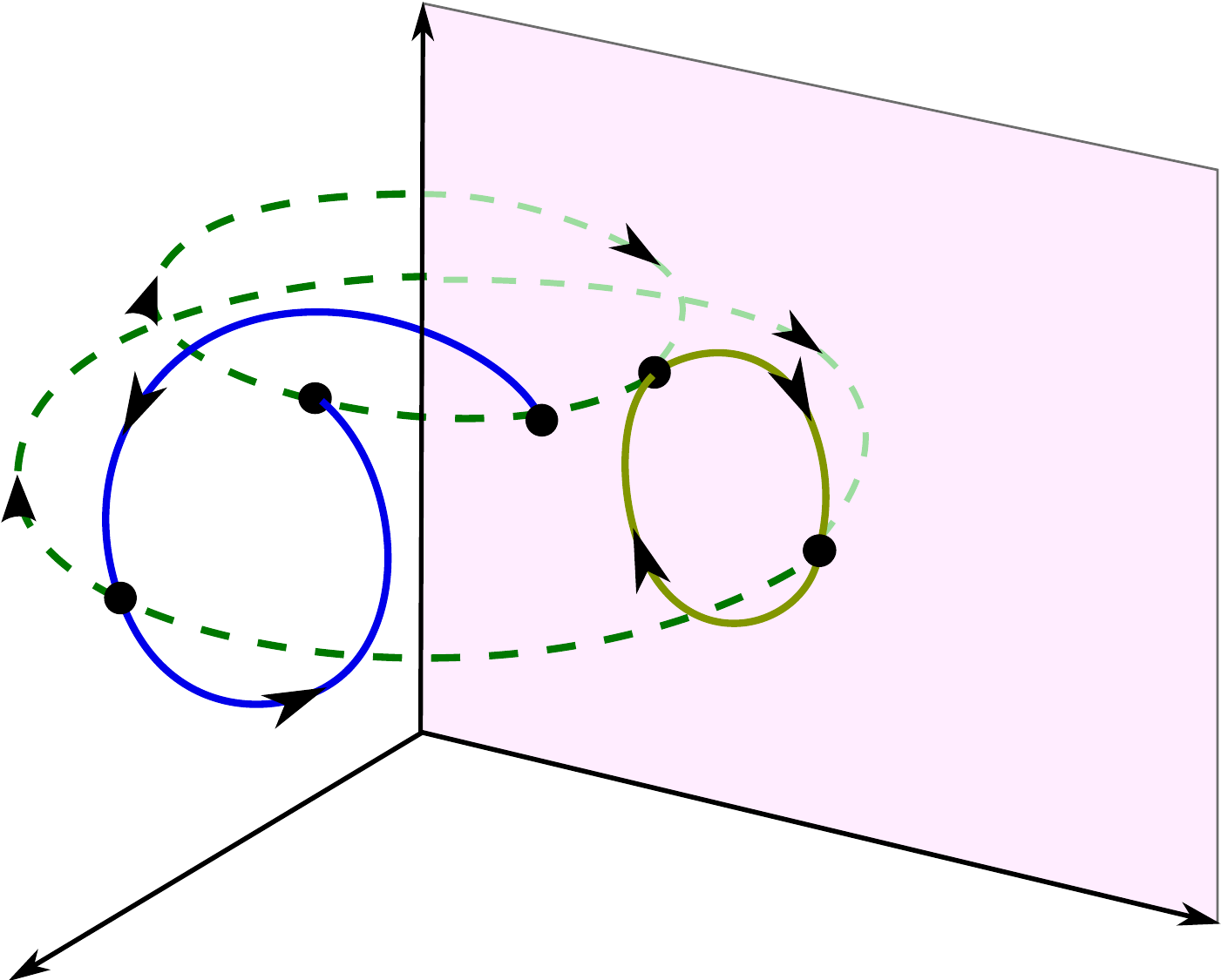}}%
    \put(0.82835153,0.19007656){\color[rgb]{0,0,0}\rotatebox{-14.84025432}{\makebox(0,0)[lb]{$\pSRed$}}}%
    \put(0.38925459,0.38713857){\color[rgb]{0,0,0}\rotatebox{0.0313674}{\makebox(0,0)[lb]{\smash{$\ssp(0)$}}}}%
    \put(0.70354118,0.30765314){\color[rgb]{0,0,0}\rotatebox{0.0313674}{\makebox(0,0)[lb]{\smash{$\sspRed(\zeit)$}}}}%
    \put(0.00068739,0.2359574){\color[rgb]{0,0,0}\rotatebox{0.0313674}{\makebox(0,0)[lb]{\smash{$\ssp(\zeit)$}}}}%
    \put(0.12576193,0.3969256){\color[rgb]{0,0,0}\rotatebox{0.0313674}{\makebox(0,0)[lb]{\smash{$\ssp(\period{p})$}}}}%
    \put(0.54113911,0.43476963){\color[rgb]{0,0,0}\rotatebox{0.0313674}{\makebox(0,0)[lb]{\smash{$\sspRed(0)$}}}}%
  \end{picture}%
 \end{center}
 \caption{
The \mslices, a \statesp\ visualization:
(a)
A chart $\pSRed \subset \pS/\Group$ lies in the $(d\!-\!N)$\dmn\
\slice\ hyperplane \refeq{PCsectQ0} normal to $\sliceTan{1}...\sliceTan{N}$, which
span the $N$\dmn\ space tangent to the group orbit $\LieEl\,\slicep$
(dotted line) evaluated at the {\template} point $\slicep$. The
hyperplane intersects {all} full \statesp\ group orbits (green
dashes).  The full \statesp\
trajectory $\ssp(\zeit)$ (blue) and the \reducedsp\ trajectory
$\sspRed(\zeit)$ (green) are equivalent up to a `moving frame' rotation
$\ssp(\zeit)=\LieEl(\zeit)\,\sspRed(\zeit)$, where $\LieEl(\zeit)$ is a
shorthand for $\LieEl(\gSpace(\zeit))$.
(b)
In the full \statesp, a \rpo\ $\ssp(0) \to \ssp(\zeit) \to
\ssp(\period{p})$ returns to the group orbit of $\ssp(0)$ after a time
$\period{p}$,  such that $\ssp(0)=\LieEl _p  \ssp
(\period{p})$. A generic \rpo\ quasi-\-periodically fills out what is
topologically a torus (\reffig{fig:CLf01group}\,(b)). In the \slice,
the symmetry-reduced trajectory is periodic, $\sspRed(0) =
\sspRed(\period{p})$.
 }\label{fig:slice}
 \end{figure}

\begin{figure}
 \begin{center}
  \setlength{\unitlength}{0.20\textwidth}
(a)
  \begin{picture}(1,0.91596465)%
    \put(0,0){\includegraphics[width=\unitlength]{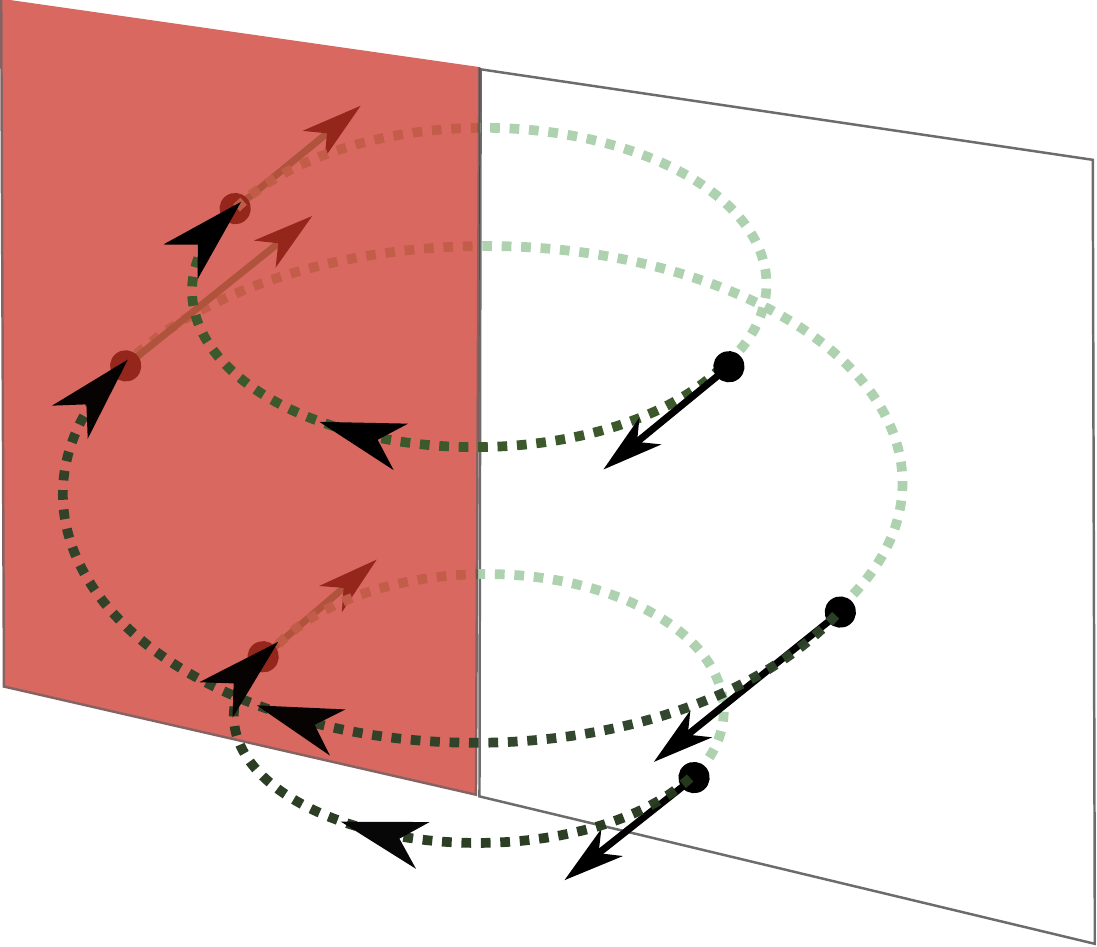}}%
    \put(0.84045332,0.62950567){\color[rgb]{0,0,0}\makebox(0,0)[lb]{\smash{$\pSRed$}}}%
    \put(0.08835189,0.08720037){\color[rgb]{0,0,0}\makebox(0,0)[lb]{\smash{$\LieEl\slicep$}}}%
    \put(0.67299795,0.12547195){\color[rgb]{0,0,0}\makebox(0,0)[lb]{\smash{$\slicep$}}}%
    \put(0.44341875,0.00071734){\color[rgb]{0,0,0}\makebox(0,0)[lb]{\smash{$\sliceTan{}$}}}%
  \end{picture}%
~~(b)\!\!
  \begin{picture}(1,0.91727402)%
    \put(0,0){\includegraphics[width=\unitlength]{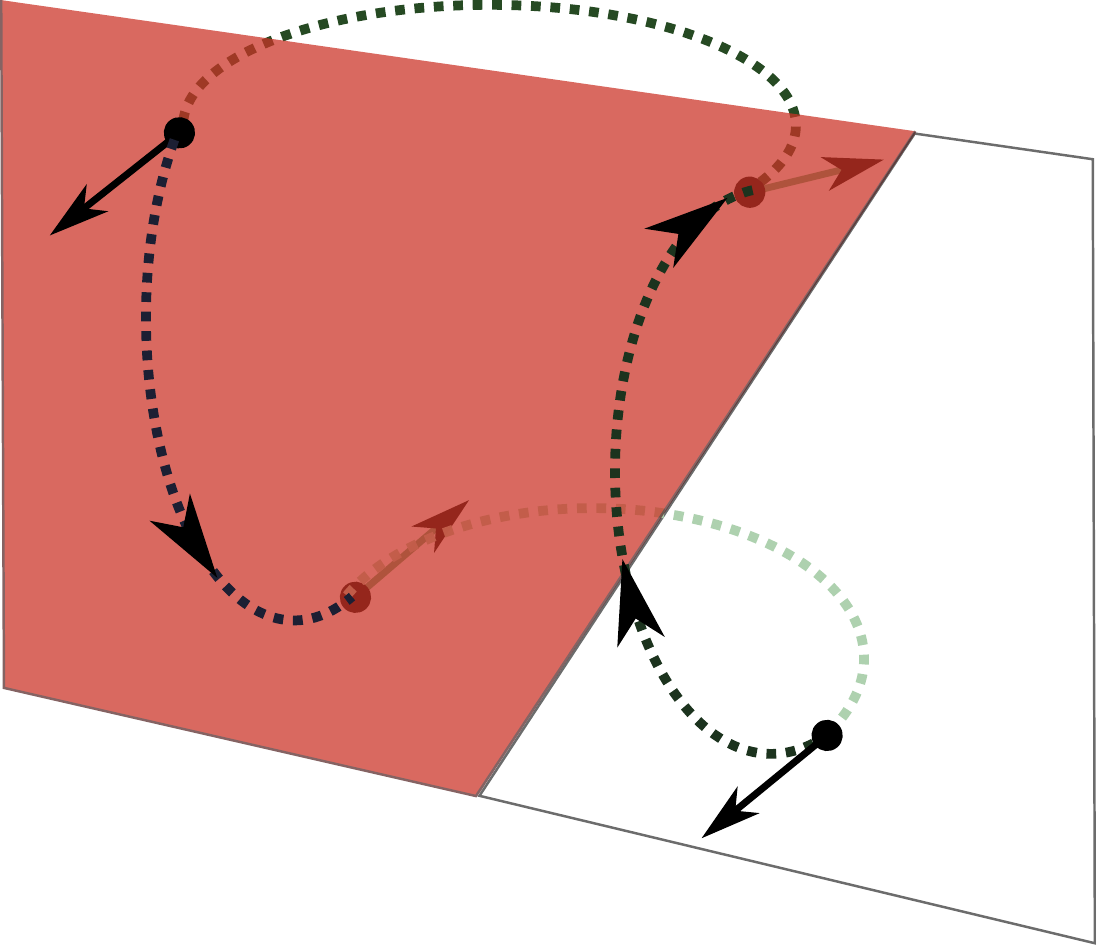}}%
    \put(0.82257887,0.59549577){\color[rgb]{0,0,0}\makebox(0,0)[lb]{\smash{$\pSRed$}}}%
    \put(0.80526889,0.1997715){\color[rgb]{0,0,0}\makebox(0,0)[lb]{\smash{$\slicep$}}}%
    \put(0.55844296,0.0631667){\color[rgb]{0,0,0}\makebox(0,0)[lb]{\smash{$\sliceTan{}$}}}%
    \put(0.61811177,0.33705605){\color[rgb]{0,0,0}\makebox(0,0)[lb]{\smash{$\LieEl\slicep$}}}%
  \end{picture}%
 \end{center}
 \caption{\label{fig:chartBord}  
The \chartBord\ ${\cal}$ is the {$(d\!-\!N\!-1)$\dmn\ } hyperplane that contains
all the points $\sspRSing$ whose group tangents $\groupTan(\sspRSing)$
lie in the \slice\ hyperplane or vanish and are thus normal to
$\sliceTan{}$. Beyond this boundary, the group orbits pierce the \slice\
hyperplane in the wrong direction, so \emph{only} the half-hyperplane
that contains the \template\ belongs to the slice. The {\chartBord} is
not easy to visualize; For the lack of dimensions, here it is drawn as a
`line', the $z$ axis in this 3\dmn\ sketch. (a) If the equivariant
coordinates transform only under the $m=1$ representation of $\SOn{2}$,
every group orbit is a circle, and crosses any slice hyperplane exactly
twice. However, if there are coordinates that transform as higher $m$,
the group orbit can pierce the hyperplane up to $2m$ times, and the
{\chartBord} lies closer to the template:  For example, (b) a group orbit
for a combination of $m=1$ and $m=2$ equivariant coordinates resembles
the seam of a baseball, and can cross the \emph{slice hyperplane} 4 times,
out of which only the point closest to the \template\ is in the
\emph{slice}.
 }%
 \end{figure}

\begin{figure}
   \begin{center}
    \setlength{\unitlength}{0.35\textwidth}
  \begin{picture}(1,0.65502164)%
    \put(0,0){\includegraphics[width=\unitlength]{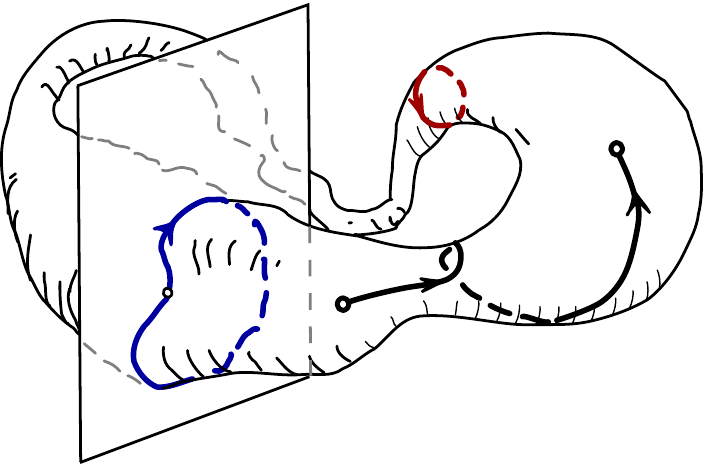}}%
    \put(0.56885811,0.58061626){\color[rgb]{0,0,0}\makebox(0,0)[lb]{\smash{$\hat{p}$}}}%
    \put(0.19355703,0.36522115){\color[rgb]{0,0,0}\makebox(0,0)[lb]{\smash{$\hat{p}$}}}%
    \put(0.13058096,0.24546158){\color[rgb]{0,0,0}\makebox(0,0)[lb]{\smash{$\sspRed(0)$}}}%
    \put(0.45027293,0.26341945){\color[rgb]{0,0,0}\makebox(0,0)[lb]{\smash{$\ssp(0)$}}}%
    \put(0.81743956,0.47684048){\color[rgb]{0,0,0}\makebox(0,0)[lb]{\smash{$\ssp(T_p)$}}}%
  \end{picture}%
  \end{center}
   \caption{\label{fig:sliceimage}
\Wurst, sliced.
      Every \slice\ hyperplane cuts every group orbit at least twice (see
      \reffig{fig:slice}), once at the orbit's closest passage to the
      {\template}, and another time at the most distant passage, also
      satisfying the \slice\ condition \refeq{PCsectQ0}. An $\SOn{2}$ \rpo\
      $\pS_p$ is topologically a torus, so the two cuts are the two \po\ images
      of the same \rpo, the good close one $\sspRed_p$ (blue), and the bad distant one (red), on
      the other side of \chartBord, and thus not in the \slice.
   }
\end{figure}

The minimal distance satisfies
the extremum condition (\reffig{fig:BeThMovFr}\,(d))
\[
\frac{\partial}{\partial \gSpace} \Norm{\ssp - \LieEl(\gSpace)\,\slicep}^2
   =
2\, \braket{\sspRed - \slicep}{\sliceTan{}}
   = 0
        \,,\quad
\sliceTan{} = \Lg \slicep
\,,
\]
where the $[d\!\times\!d]$ matrix $\Lg$ is the generator of infinitesimal
symmetry transformations. $\Norm{\LieEl(\gSpace)\slicep}= \Norm{\slicep}$
is a constant.
{
To streamline the exposition, we shall assume here that the symmetry
group is $\SOn{n}$. In that case $\Lg$ is antisymmetric, so the group
tangent vector $\sliceTan{}$ evaluated at $\slicep$ is normal to
$\slicep$ and the term $\braket{\slicep}{\Lg\,\slicep}$ vanishes.}
Therefore  $\sspRed$, the point on the group orbit of $\ssp$ that lands
in the \slice\ satisfies the \emph{\slice\ condition}
\beq
\braket{\sspRed}{\sliceTan{}} = 0
    \,.
\ee{PCsectQ0}
As $\ssp(\zeit)$ varies in time, the {\template} $\slicep$ tracks the
motion using the \slice\ condition \refeq{PCsectQ0} to minimize
$\Norm{\ssp(\zeit)-\LieEl(\phi(\zeit))\slicep}$, and the full-space
trajectory $\ssp(\zeit)$ is rotated into the {\reducedsp} trajectory
$\sspRed(\zeit)$ by appropriate time varying \emph{moving frame} angles
$\{\gSpace(\zeit)\}$, as depicted in \reffig{fig:slice}\,{(a)}. $\pSRed$
is thus a $(d\!-\!N)$\dmn\ hyperplane normal to the $N$ group tangents
evaluated at the \slicep\ as sketched in \reffig{fig:slice} in a highly
idealized manner: A group orbit is an $N$\dmn\ manifold and, even for
$\SOn{2}$, is usually only topologically a circle and can intersect a
hyperplane any number of times  (see
\reffigs{fig:chartBord}{fig:sliceimage}).


One can write the equations for the flow in the \reducedsp\
$\dot{\sspRed} = \velRed(\sspRed)$ (for details see, for example,
\refref{DasBuch}) as
\bea
\velRed(\sspRed) &=& \vel(\sspRed)
     \,-\, \dot{\gSpace}(\sspRed) \, \groupTan(\sspRed)
\label{EqMotMFrame}\\
\dot{\gSpace}(\sspRed) &=& \braket{\vel(\sspRed)}{\sliceTan{}}
                       /\braket{\groupTan(\sspRed)}{\sliceTan{}}
\,,
\label{reconstrEq}
\eea
which confines the motion to the \slice\ hyperplane. Thus, the dynamical
system $\{\pS,\map^t\}$ with continuous symmetry \Group\ is replaced by
the {\reducedsp} dynamics $\{\pSRed,\mapRed^t\}$: The velocity in the
full \statesp\ $\vel$ is the sum of $\velRed$, the velocity component in
the \slice\ hyperplane, and $\dot{\gSpace}\,\groupTan$, the velocity
component along the group tangent space. The integral of the {\em
reconstruction equation} for $\dot{\gSpace}$ keeps track of the group
shift in the full \statesp.

The {\template} $\slicep$ should be a generic \statesp\ point in the
sense that its group orbit has the full $N$ dimensions of the group
\Group. The set of the group orbit points {closest} to the \template\
\slicep\ forms a neighborhood of \slicep\ in which each group orbit
intersects the hyperplane \emph{only once}. A \slice\ hyperplane
qualitatively captures neighboring group orbits until, for a point
$\sspRSing$ not so close to the \template, the group tangent vector
$t(\sspRSing)$ lies in the \slice\ hyperplane. The group orbits for such
points are grazed tangentially rather than sliced transversally, much
like what happens at the \poincBord\ \refeq{eq:sspRSing} for evolution in
time. This is also a linear condition and defines the \chartBord\ ${\cal
S}$,\rf{SiCvi10,FrCv11} a $(d\!-\!N\!-1)$\dmn\ manifold, which contains
all the points $\sspRSing$ whose group tangents lie in the \slice\
hyperplane, \ie,
\beq
\braket{\sspRSing}{\sliceTan{}} \,=\, 0
      \mbox{ and }
\braket{\groupTan(\sspRSing)}{\sliceTan{}} \,=\, 0
\,.
\label{sliceSingl0}
\eeq
${\cal S}$ also contains all points for which $\groupTan(\sspRSing)=0$.
While for the \PoincSec s \refeq{eq:sspRSing} the analogous points were
\eqva\ (captured only if the section cut through them), for \slice\
hyperplanes points with vanishing group actions belong to invariant
subspaces, and, by its definition, every {\chartBord} automatically
includes \emph{all} invariant subspaces.

For the \cLe\ \refeq{eq:CLeR}, the invariant subspace is the 1\dmn\
$z$-axis, with trivial dynamics, $z=-bz$, but in general invariant
subspaces are high\dmn\ and have their own dynamics.
{
Physicists, for example general relativists, often work in invariant
subspaces, as this is easier than solving the full problem.\rf{SCD07}
Such approaches yield highly symmetric solutions,\rf{SKMacHH09,HGC08}
whose dynamics may be quite different from those that guide turbulence in
the full \statesp\ (for a striking example, see \refref{ACHKW11}).}

There is yet another, much kinder type of a border: a \emph{ridge}. Our
initial chart $\pSRed{}^{(1)}$ is a {($d\!-\!N$)\dmn\ } hyperplane. If we
pick another {\template} point $\slicep{}^{(2)}$, it comes along with its
own \slice\ hyperplane $\pSRed{}^{(2)}$. Any pair of {$(d\!-\!N)$\dmn\ }
local \slice\ hyperplanes intersects in a {ridge}, a {$(d\!-\!N-\!1)$\dmn\ }
hyperplane {\PoincS} of points $\sspRed^*$ shared by a pair of charts and
thus satisfying the \slice\ condition \refeq{PCsectQ0} for both,
\beq
\braket{\sspRed^*}{\sliceTan{}{}^{(1)}} = 0
\mbox{ and }
\braket{\sspRed^*}{\sliceTan{}{}^{(2)}} = 0
    \,.
\ee{ridge}
The ridge forms a \PoincSec\ $\PoincS{}^{(ij)}$ that serves as
a toll bridge, crossed by any direct transit from a chart $\pSRed{}^{(j)}$ to the
adjacent chart $\pSRed{}^{(i)}$.
In \reffig{fig:A29-2slices}\,(a) a ridge is visualized as a `line', and
in \reffig{fig:A29-1ridge} as a `plane' of intersection of two volumes.
We shall refer to the neighborhood of a \template\ $\slicep{}^{(j)}$
bounded by its {\chartBord} and the ridges to other such linear
neighborhoods as a \emph{chart} $\pSRed{}^{(j)} \subset \pS/\Group$, and
to \refeq{sliceSingl0} and \refeq{ridge} as the {border conditions}.

\section{Charting the \slice}
\label{s:chart}

Let us summarize the voyage so far: we are charting a curved manifold,
and it would be nice to use tools of differential geometry, but this seems
not possible in the high\dmn\ \statesp\ of hydrodynamics turbulence.
The only feasible way to chart this space is to (1) quotient all continuous
symmetries, and (2) tile the reduced \statesp\ with flat {$(d\!-\!N)$\dmn\ } tiles,
or charts. We do this step by step, starting with a set of \template s and using
them to construct charts of each neighborhood, and then building up an atlas
of the \emph{\slice}, chart by chart, which captures all of the reduced
dynamics of interest (but not all possible dynamics). Here are the steps along
the way:

\begin{description}

\item[\Template]
Pick a {\template} $\slicep$ such that $\Group$ acts on it
regularly with a group orbit of dimension $N$.

\item[\Slice\ hyperplane]
The $(d\!-\!N)$\dmn\ hyperplane satisfying
\( 
\braket{\sspRed}{\sliceTan{a}}=0
\,,
\) 
normal to group transformation directions at the {\template} $\slicep$.
\index{slice}

\item[Moving frame]
For any $\ssp$, the \slice\ condition $\braket{\sspRed}{\sliceTan{}}=0$
on $\ssp = \LieEl(\gSpace)\sspRed$ determines the moving frame, \ie, the
group action $\LieEl(\gSpace)$ that brings $ \ssp$ into the \slice\
hyperplane.
\index{moving frame}

\item[\ChartBord]
The set of points $\sspRSing$ on a \slice\ hyperplane whose group orbits
graze the hyperplane tangentially, such that
$\braket{\sspRSing}{\sliceTan{}} \,=\,
\braket{\groupTan(\sspRSing)}{\sliceTan{}} \,=\, 0 \,.$

\item[Flow invariant subspace]
If a subset or all of the group tangents of a \chartBord\ point
$\sspRSing$ vanish, $\groupTan_a(\sspRSing)=0$, its time  trajectory
remains within a flow-invariant subspace for all times.

\item[Ridge]
A hyperplane of points $\sspRed^* \in \PoincS{}^{(21)}$ formed by the
intersection of a pair of \slice\ hyperplanes $\pSRed{}^{(1)}$ and
$\pSRed{}^{(2)}$.

\item[Chart]
The neighborhood of a \template\ $\slicep{}^{(j)}$, bounded by the
{\chartBord} and the ridges to other linear neighborhoods, comprises
a \emph{chart} $\pSRed{}^{(j)} \subset \pS/\Group$. The
borders ensure that there is no more than one oriented group orbit traversal per chart;
a group orbit either pierces one chart, or no charts at all.

\item[Atlas]
A set of $(d\!-\!N)$\dmn\ contiguous charts $\pSRed{}^{(1)},
\pSRed{}^{(2)}, \cdots$

\item[\Slice]
Let $\Group$ act on a $d$\dmn\ manifold $\pS$, with group
orbits of dimension $N$ or less. A \emph{\slice} is a $(d\!-\!N)$\dmn\
submanifold $\pSRed$ such that all group orbits that
intersect $\pSRed$ do so transversally and only once.

\end{description}

 \begin{figure}
 \begin{center}
  \setlength{\unitlength}{0.30\textwidth}
(a)\;\;
  \begin{picture}(1,0.92174023)%
    \put(0,0){\includegraphics[width=\unitlength]{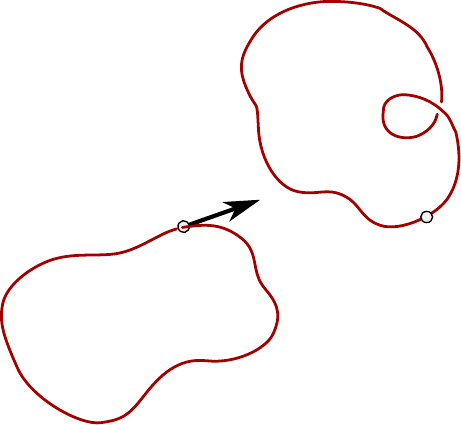}}%
    \put(0.38186388,0.34995272){\color[rgb]{0,0,0}\makebox(0,0)[lb]{\smash{$\slicep{}^{(1)}$}}}%
    \put(0.41769945,0.50079738){\color[rgb]{0,0,0}\makebox(0,0)[lb]{\smash{$\sliceTan{}{}^{(1)}$}}}%
    \put(0.87339467,0.35886318){\color[rgb]{0,0,0}\makebox(0,0)[lb]{\smash{$\ssp'{}^{(2)}$}}}%
  \end{picture}%
\\
(b)\;\;
  \begin{picture}(1,1.14107266)%
    \put(0,0){\includegraphics[width=\unitlength]{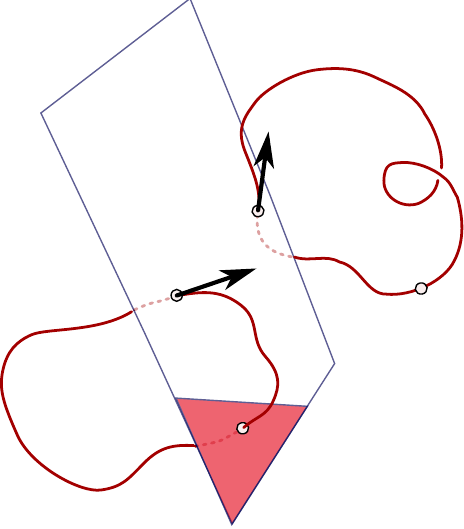}}%
    \put(0.54442322,0.48851386){\color[rgb]{0,0,0}\makebox(0,0)[lb]{\smash{$\sliceTan{}{}^{(1)}$}}}%
    \put(0.33649368,0.41474621){\color[rgb]{0,0,0}\makebox(0,0)[lb]{\smash{$\slicep{}^{(1)}$}}}%
    \put(0.41644523,0.62981493){\color[rgb]{0,0,0}\makebox(0,0)[lb]{\smash{$\slicep{}^{(2)}$}}}%
    \put(0.616644,0.84680609){\color[rgb]{0,0,0}\makebox(0,0)[lb]{\smash{$\sliceTan{}{}^{(2)}$}}}%
    \put(0.88017597,0.4261647){\color[rgb]{0,0,0}\makebox(0,0)[lb]{\smash{$\ssp'{}^{(2)}$}}}%
    \put(0.29194797,0.95658666){\color[rgb]{0,0,0}\makebox(0,0)[lb]{\smash{$\pSRed{}^{(1)}$}}}%
  \end{picture}%
 \end{center}
 \caption{\label{fig:A29-2tmplts}
A 2-chart atlas. Sketch
    (a)
depicts two templates $\slicep{}^{(1)}$, $\ssp'{}^{(2)}$, each with its
group orbit. Start with the {\template} $\slicep{}^{(1)}$. All group
orbits traverse its $(d\!-\!1)$\dmn\ \slice\ hyperplane, including the
group orbit of the second {\template} $\ssp'{}^{(2)}$.
    (b)
Replace the second {\template} by its closest group orbit point
$\slicep{}^{(2)}$, \ie, the point in chart $\pSRed{}^{(1)}$. This is
allowed as long as  $\slicep{}^{(2)}$ is closer than the $\pSRed{}^{(1)}$
{\chartBord} (red region), otherwise an interpolating, closer template
needs to be introduced.
 }
 \end{figure}

 \begin{figure}
 \begin{center}
  \setlength{\unitlength}{0.40\textwidth}
(a)\;\;
  \begin{picture}(1,0.86567815)%
    \put(0,0){\includegraphics[width=\unitlength]{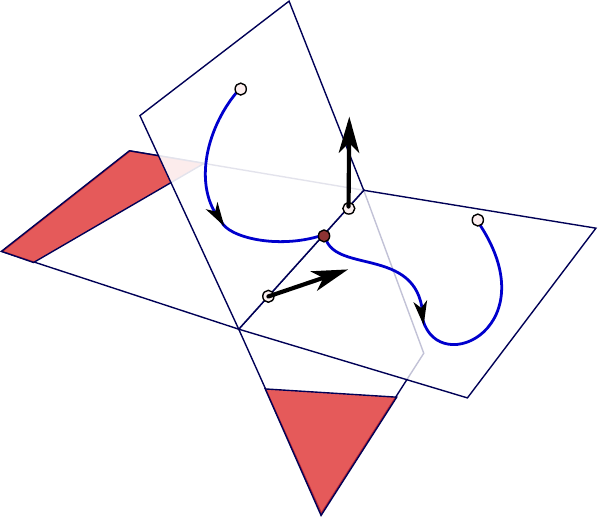}}%
    \put(0.3850416,0.38725438){\color[rgb]{0,0,0}\makebox(0,0)[lb]{\smash{$\slicep{}^{(1)}$}}}%
    \put(0.60194012,0.48012421){\color[rgb]{0,0,0}\makebox(0,0)[lb]{\smash{$\slicep{}^{(2)}$}}}%
    \put(0.4042968,0.74412842){\color[rgb]{0,0,0}\makebox(0,0)[lb]{\smash{$\sspRed(0)$}}}%
    \put(0.79647438,0.54627847){\color[rgb]{0,0,0}\makebox(0,0)[lb]{\smash{$\sspRed(\zeit)$}}}%
  \end{picture}%
\\
(b)\;\;
  \begin{picture}(1,0.5127804)%
    \put(0,0){\includegraphics[width=\unitlength]{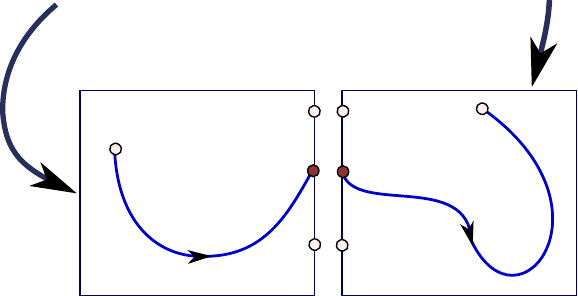}}%
    \put(0.16199231,0.03841546){\color[rgb]{0,0,0}\makebox(0,0)[lb]{\smash{$\pSRed{}^{(1)}$}}}%
    \put(0.63051777,0.0374085){\color[rgb]{0,0,0}\makebox(0,0)[lb]{\smash{$\pSRed{}^{(2)}$}}}%
    \put(0.21517269,0.28787637){\color[rgb]{0,0,0}\makebox(0,0)[lb]{\smash{$\sspRed(0)$}}}%
    \put(0.75921701,0.25014044){\color[rgb]{0,0,0}\makebox(0,0)[lb]{\smash{$\sspRed(\zeit)$}}}%
    \put(0.60952792,0.26511997){\color[rgb]{0,0,0}\makebox(0,0)[lb]{\smash{$\slicep{}^{(2)}$}}}%
    \put(0.45827029,0.02997228){\color[rgb]{0,0,0}\makebox(0,0)[lb]{\smash{$\slicep{}^{(1)}$}}}%
  \end{picture}%
 \end{center}
 \caption{\label{fig:A29-2slices}
A 2-chart atlas.
    (a)
Now that the group orbits have been reduced to points, erase them and
consider the two \slice\ hyperplanes through the two {\template s}. As
these two {\template s} are the closest points viewed from either group
orbit, they lie in both \slice\ hyperplanes. However, the two tangent
vectors $\sliceTan{}{}^{(1)}$ and $\sliceTan{}{}^{(2)}$ have different
orientations, so they define two distinct charts $\pSRed{}^{(1)}$ and
$\pSRed{}^{(2)}$ which intersect in the \emph{ridge}, a hyperplane of
dimension $(d\!-\!2)$ (here drawn as a `line', and in
\reffig{fig:A29-1ridge} as intersection of two `volumes') shared by the
template pair that satisfies both \slice\ conditions \refeq{ridge}. The
chart for the neighborhood of each template (a page of the atlas in part
(b)) extends only as far as this ridge. If the templates are sufficiently
close, the {\chartBord} of each \slice\ hyperplane (red region) is beyond
this ridge, and not encountered by the symmetry-reduced trajectory
$\sspRed(\zeit)$. The reduced trajectory is continuous in the slice
comprised of such charts - it switches the chart whenever it crosses a
ridge.
    (b)
The slice (unique point for each group orbit) is now covered by an atlas
consisting of $(d\!-\!1)$\dmn\ charts $\pSRed{}^{(1)}, \pSRed{}^{(2)},
\cdots$.
 }
 \end{figure}

In the literature,\rf{Mostow57,Pal61,GuiSte90} `\slice' refers to any
co-dimension~$N$ manifold that slices transversally a group orbit. Here, we define
an atlas over a \slice\ constructively but more narrowly, as a contiguous
set of flat charts, with every group orbit accounted for by the
atlas sliced only once, and belonging to a single chart. A \slice\ is not
global, it slices only the group orbits in an open neighborhood of the
\statesp\ region of interest.

The physical task, for a given dynamical flow, is to pick a set of
qualitatively distinct {\template s} (for a turbulent pipe
flow there might be one typical of 2-roll states, one for 4-roll states, and so on),
which together provide a good atlas for the region of $\pS/\Group$
explored by chaotic trajectories.

The rest is geometry of hyperplanes and has nothing to do with dynamics.
Group orbits $\pS_{\ssp^{(j)}}$ through $\ssp^{(j)}$, group tangents
$\groupTan(\slicep{}^{(j)})$, and the associated charts $\pSRed{}^{(j)}$
are purely group-theoretic concepts. The \slice, \chartBord\ and ridge
conditions \refeq{PCsectQ0}, \refeq{sliceSingl0} and \refeq{ridge} are
all linear conditions which depend on the ray defined by the \template\
\slicep, not its magnitude. Checking whether the {\chartBord} is on the
far side of the ridge between two \slice\ hyperplanes is a linear
computation; for a symmetry-reduced trajectory moving in $\pSRed{}^{(1)}$
chart one only has to keep checking the sign of
\beq
\braket{\sspRed(\zeit)}{\sliceTan{}{}^{(2)}}
\,.
\ee{eq:chartBord}
Once the sign changes, the ridge has been crossed, and from then on
the trajectory should be reduced to the $\pSRed{}^{(2)}$ chart.
For three or more charts you will have to align the ridge of the current
chart with a previously-used chart. You'll cross that ridge when you come
to it (a hint: the manifold is curved, so there will be a finite jump in
phase).

\begin{figure}
 \begin{center}
  \setlength{\unitlength}{0.30\textwidth}
  \begin{picture}(1,0.89907101)%
    \put(0,0){\includegraphics[width=\unitlength]{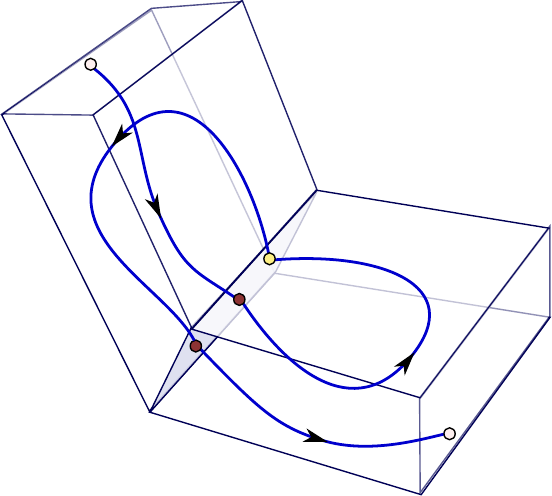}}%
    \put(0.00894598,0.81885604){\color[rgb]{0,0,0}\makebox(0,0)[lb]{\smash{$\sspRed(0)$}}}%
    \put(0.88743345,0.05105926){\color[rgb]{0,0,0}\makebox(0,0)[lb]{\smash{$\sspRed(\zeit)$}}}%
    \put(0.78614059,0.43443027){\color[rgb]{0,0,0}\rotatebox{-25.76142111}{\makebox(0,0)[lb]{\smash{$\pSRed{}^{(2)}$}}}}%
    \put(0.37048948,0.79485578){\color[rgb]{0,0,0}\rotatebox{-61.41291822}{\makebox(0,0)[lb]{\smash{$\pSRed{}^{(1)}$}}}}%
    \put(0.2429401,0.27697318){\color[rgb]{0,0,0}\makebox(0,0)[lb]{\smash{$\sspRed_2$}}}%
    \put(0.47832196,0.33514069){\color[rgb]{0,0,0}\makebox(0,0)[lb]{\smash{$\sspRed_1$}}}%
  \end{picture}%
 \end{center}
 \caption{\label{fig:A29-1ridge}
Here the two charts of \reffig{fig:A29-2slices}\,(a) are drawn as
two $(d\!-\!1)$\dmn\ slabs. The ridge, their $(d\!-\!2)$\dmn\ intersection,
can then be drawn as the shaded plane. This hyperplane cuts across the
symmetry-reduced trajectory $\sspRed(\zeit)$ and thus serves as a
\PoincSec\ $\PoincS{}^{(21)}$ that captures all transits from the
neighborhood of {\template} $\slicep{}^{(1)}$ to the neighborhood of
{\template} $\slicep{}^{(2)}$. \PoincSec\ transits are oriented, so
$\sspRed_1$ and $\sspRed_2$ are in the section, but the third point is not.
 }
 \end{figure}

How the charts are put together is best told as a graphic tale, in the 5
frames of Figs.~\ref{fig:A29-2tmplts}, \ref{fig:A29-2slices} and
\ref{fig:A29-1ridge}, and then illustrated by contrasting the mess of the
\cLe\ strange attractor \reffig{fig:CLf01group}\,(a) to the elegance of
its 2-chart atlas, \reffig{fig:CLf01group}\,(d).

It is worthwhile to note that the only object that enters the \slice\
hyperplane, border and ridge conditions is the ray defined by the unit
vector $\hat{t}{}^{'}= \sliceTan{}/\Norm{\sliceTan{}}$. This gives much
freedom in picking \template s. In particular, the two rays
\bea
\hat{t}{}^{'(1)} &=& (0.263,-0.692,0.624,-0.251,0)
    \continue
\hat{t}{}^{'(2)} &=& (0.153, -0.610, 0.747, -0.213, 0)
\label{DanielTmpls2}
\eea
used to construct the \cLe\ 2-chart atlas of
\reffig{fig:CLf01group}\,(d) were found by numerical experimentation.

With the atlas in hand, the dynamics is fully charted: as explained in
\refrefs{DasBuch,SiCvi10}, Poincar\'e return maps then yield all
admissible \rpo s.

Three concluding remarks on what \slice s \emph{are not}:

(1) Symmetry reduction is not a dimensional-reduction scheme, a
projection onto fewer coordinates, or flow modeling by fewer degrees of
freedom: It is a local change of coordinates with one (or $N$) coordinate(s)
pointing along the continuous symmetry directions. No information is lost
by symmetry `reduction', one can go freely between solutions in the full
and reduced \statesp s by integrating the associated {reconstruction
equations} \refeq{reconstrEq}.

(2) If the flow is also invariant under discrete symmetries, these should
be reduced by methods described, for example, in ChaosBook.org.

(3) An atlas is \emph{not needed} for Newton determination of a single
invariant solution, or a study of its bifurcations.\rf{golubII} Any local
section and slice plus time and shift constraints does the job.%
\rf{Visw07b,duguet08,mellibovsky11} It is
possible to compute 60,000 \rpo s this way.\rf{SCD07} Once we have more
than one invariant solution, the question is: how is this totality of
solutions interrelated? For that, a good atlas is a necessity.

\section{Bridges to nowhere}
\label{s:bridge}

Everybody encounters a symmetry sooner or later, so the literature on
symmetry reduction is vast (for a historical overview, see remarks in
ChaosBook.org and \refref{SiCvi10}). Before asking, ``Why the \mslices\
and not [...]?'' a brief tour of the more familiar symmetry reduction
schemes is called for. They all have one thing in common: they will not
work for high\dmn\ nonlinear systems.

To start with, mastery of quantum-mechanics or bifurcation
theory\rf{ruell73,golubII} symmetry reduction to linear irreducible
representations is only partially illuminating; linear theory
works quite well for linear unitary operators or close to a bifurcation,
but, as we tried to show in this pictorial tour, the way symmetries act
on nonlinear systems is much subtler. For flows with strongly nonlinearly
coupled modes, both time trajectories and  group orbits are complicated,
so choices of sections and \slice s require insight into the geometry of
the particular flow, there exists no general theory of linear
transformations into symmetry irreducible coordinates that would do the
job.

There are purely group-theoretical approaches, with no dynamics to inform
them, inspired by the observation that while coordinates $x_i$ are
equivariant, the squared length $r^2 = \sum x_i^2$ is \emph{invariant}
under $\On{n}$ transformations.

For $\SOn{2}$, an obvious idea is to go to polar coordinates. The
simplest nonlinear examples\rf{AGHO288} already run into $r_j \to 0$ type
of singularities, and it is not altogether clear how one would rewrite
the \NSe\ in such a format, or integrate them numerically. A more
sophisticated  approach is to rewrite the dynamics in terms of
invariant polynomial bases, described lucidly in \refref{GL-Gil07b}, with
the equivariant \statesp\ coordinates $(x_1,x_2,x_3,...,x_d)$ replaced by
an invariant polynomial basis $(u_1,u_2,u_3,...,u_m)$. As the dimension of
the problem increases, the number of these polynomials grows
quickly, as does the number of syzygies, the nonlinear relations amongst
them. There is no guiding principle for picking a set of such
polynomials, and no practical way to implement the
scheme\rf{gatermannHab} for high\dmn\ flows: how and why would one
replace the {large number of} equivariant \statesp\ coordinates of
hydrodynamic turbulence with a vast number of invariant polynomials?

Others approaches are informed by dynamics, foremost among them being the
method of {co-moving frames}. Visualizing a single `relative' trajectory
in its co-moving frame, \ie, moving with that solution's mean
{\phaseVel}, is useful if one is concerned with that individual solution
and the tiny \rpo s (modulated-amplitude waves) that bifurcate off
it.\rf{duguet08,mellibovsky11} A co-moving frame is useless, however, if
we are concerned with studying collections of these trajectories, as each
solution travels with its own mean {\phaseVel}
{$\velRel_p = \gSpace_p/\period{p}$}, and there is no single
co-moving frame that can simultaneously reduce \emph{all} traveling
solutions. The \slice\ that we construct here is not `co-moving', but
\emph{emphatically} stationary.

There exists a beautiful theory of symplectic symmetry reduction for the
mechanics of three\dmn\ rigid bodies,\rf{MaWe74,AbrMars78} or
using Lie symmetry reduction to derive Eulerian velocity fields from
Lagrangian trajectories.\rf{MorrGree80} These approaches do not appear to
be applicable to problems considered here, and anyway, the goal is
different. Rather than to reduce a particular set of equations, we seek
to formulate a computationally straightforward and general method of
reducing any continuous symmetry, for \emph{any} high\dmn\
chaotic/turbulent flow. One should also note that `symmetry reduction' in
general relativity\rf{SKMacHH09} and Lie theory often implies restricting
one's solution space to a subspace of higher symmetry; here we always
work in the full \statesp.

\begin{figure}
   \centering
  \setlength{\unitlength}{0.20\textwidth}
(a)~~~
  \begin{picture}(1,0.98073806)%
    \put(0,0){\includegraphics[width=\unitlength]{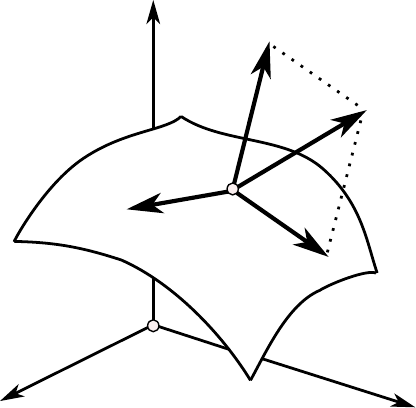}}%
    \put(0.8635648,0.73622308){\color[rgb]{0,0,0}\makebox(0,0)[lb]{\smash{$\vel$}}}%
    \put(0.49893205,0.86039365){\color[rgb]{0,0,0}\makebox(0,0)[lb]{\smash{$\vel_{\bot}$}}}%
    \put(0.27198728,0.5378933){\color[rgb]{0,0,0}\makebox(0,0)[lb]{\smash{$\groupTan_1$}}}%
    \put(0.58493215,0.33483773){\color[rgb]{0,0,0}\makebox(0,0)[lb]{\smash{$\groupTan_2$}}}%
    \put(0.51559234,0.22257862){\color[rgb]{0,0,0}\makebox(0,0)[lb]{\smash{$\pS_\ssp$}}}%
  \end{picture}%
(b)~~~
  \begin{picture}(1,0.98655417)%
    \put(0,0){\includegraphics[width=\unitlength]{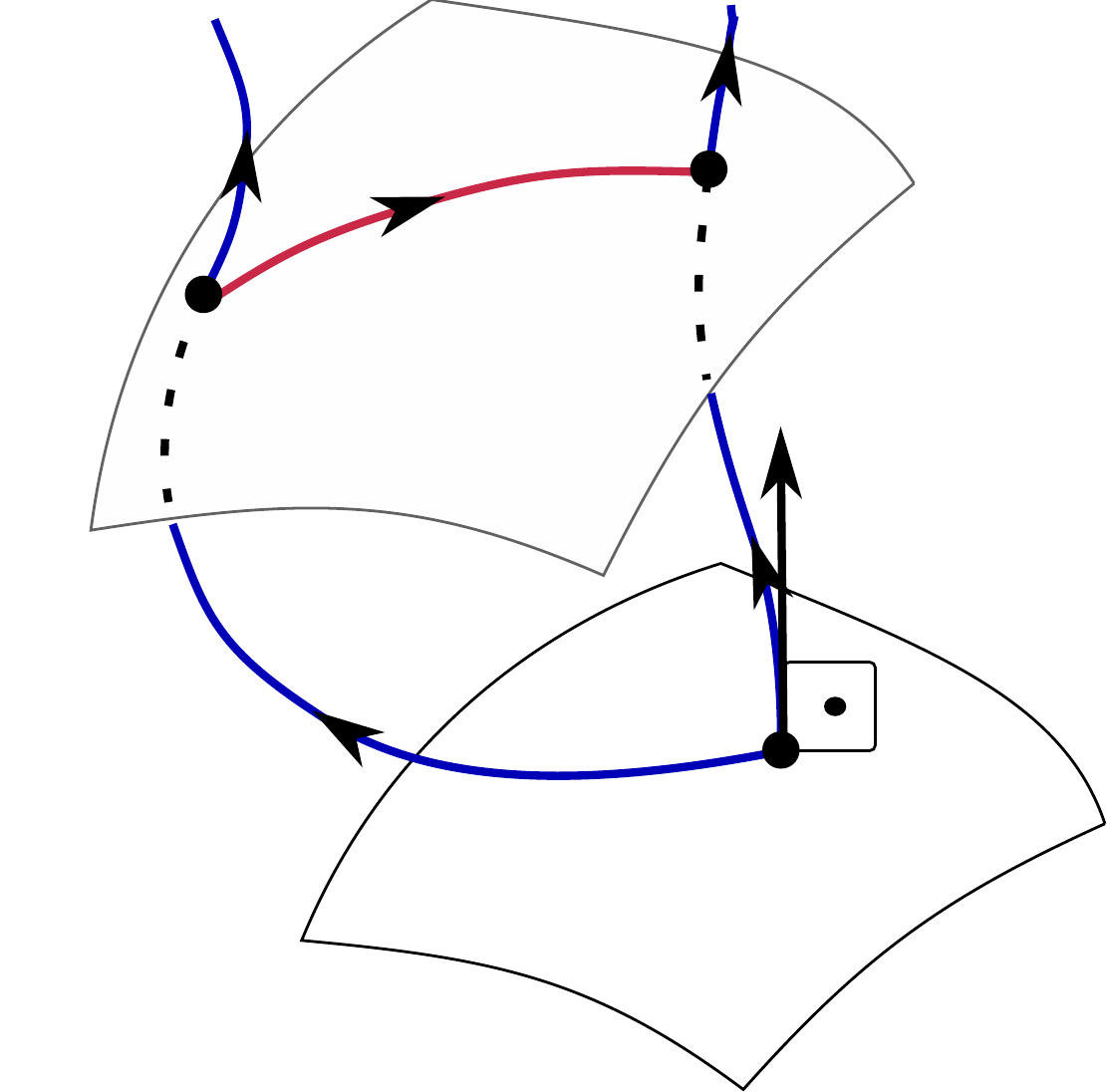}}%
    \put(0.20559239,0.64023845){\color[rgb]{0,0,0}\rotatebox{0.0313674}{\makebox(0,0)[lb]{\smash{$\ssp(\zeit)$}}}}%
    \put(0.68383186,0.20519203){\color[rgb]{0,0,0}\rotatebox{0.0313674}{\makebox(0,0)[lb]{\smash{$\ssp(0)$}}}}%
    \put(0.44475925,0.7309461){\color[rgb]{0,0,0}\rotatebox{0.0313674}{\makebox(0,0)[lb]{\smash{$\sspRed(\zeit)$}}}}%
    \put(0.31760559,0.8462057){\color[rgb]{0,0,0}\rotatebox{8.313674}{\makebox(0,0)[lb]{\smash{$\LieEl(\zeit)$}}}}%
    \put(0.70884327,0.61850672){\color[rgb]{0,0,0}\rotatebox{0.0313674}{\makebox(0,0)[lb]{\smash{$\vel_{\bot}$}}}}%
  \end{picture}%
   \caption{\label{fig:BeThMconnect}
    (a)
By equivariance $\vel(\ssp)$ can be replaced by $\vel_\bot(\ssp)$, the
velocity normal to the group tangent directions at \statesp\ point $\ssp$.
    (b)
The method of connections replaces $\vel(\sspRed)$ at every instant
$\sspRed =\sspRed(\zeit)$ by $\vel_\bot(\sspRed)$, so in
$\sspRed(\zeit)$'s covariant frame there is no motion along the group
tangent directions.
}
\end{figure}

There is, however, one intriguing, compelling  and physically informed
contender. In mechanics and field theory it is natural to separate the
flow {\em locally} into group dynamics and a transverse, `horizontal'
flow,\rf{Smale70I,AbrMars78} by the `method of
connections',\rf{rowley_reduction_2003} illustrated in
\reffig{fig:BeThMconnect}.
{
The method of connections, however, does not reduce the dynamics to a
lower\dmn\ \reducedsp\ $\pS/\Group$.
In contrast to the method of co-moving frames, where one defines a mean
{\phaseVel} of a \rpo, the method of connections is inherently local. The two
methods coincide for relative equilibria.
}

The meaning of the `method of
connections' in classical dynamics is clearest in the work of Shapere
and Wilczek:\rf{ShWi89a,ShWi06} one can observe a swimmer (or our dancer)
from a fixed \slice\ frame, or bring her back to observe only the
shape-changing dynamics, no drifting. Left to herself, she will reemerge
in the same pose someplace else: that shift is called a `geometrical
phase', which -while accruing it is the whole point of swimming- has not
played any role in our discussion of symmetry reduction. Conversely, most
gauge choices in quantum field theory are covariant, and while that
suffices to regularize path integrals, the \mslices\ says that this is no
symmetry reduction at all, and it yields no insight into the geometry of
nonlinear flows.

{
Symmetry reduction in dynamics (including classical field theories such
as the \NSe) closely parallels the reduction of gauge symmetry in
quantum field theories. There, the freedom of choosing moving frames
shown in \reffig{fig:BeThMovFr} is called `gauge freedom' and a
particular prescription for choosing a representative from each gauge
orbit is called `gauge fixing'. Just like the slice hyperplanes of
\reffig{fig:chartBord} may intersect a group orbit many times, a gauge
fixing submanifold may not intersect a gauge orbit, or it may intersect
it more than once (`Gribov ambiguity').\rf{Gribov77,VaZw12} In this
context a chart is called a `Gribov' or `fundamental modular' region and
its border is called a `Gribov horizon' (a convex manifold in the
space of gauge fields). The Gribov region is compact and bounded by the
Gribov horizon. Within a Gribov region the `Faddeev-Popov operator'
(analogue of the group orbit tangent vector) is strictly positive, while on
the Gribov horizon it has at least one vanishing eigenvalue.
}

\section{Conclusions}
\label{s:concl}

As turbulent flow evolves, every so often we catch a glimpse of a
familiar structure. For any finite spatial resolution and time, the flow
follows  unstable {\cohStr s} belonging to an alphabet of representative
states, here called `\template s'. However, in the presence of
symmetries, near recurrences can be identified only if shifted both in
time and space.

In the method of sections (along time direction) and slices (along
spatial symmetry directions), the identification of physically nearby
states is achieved by cutting group orbits with a finite set of
hyperplanes, one for each continuous parameter, with each time trajectory
and group orbit of symmetry-equivalent points represented by a single
point. The \mslices\ is akin to (but distinct
from) cutting across trajectories by means of sections. Both methods
reduce continuous symmetries: one sections the continuous-time
trajectories, the other slices the layers of the onion formed by
group-orbits. Both are triggered by analogous conditions: oriented
piercing of the section and oriented piercing of the slice. Just as a
\PoincSec\ goes bad, the \slice\ hyperplane goes bad the moment
transversality is lost. A slice, however, is emphatically \emph{not} a
\PoincSec:
as the first step in a reduction of dynamics, a slice replaces a trajectory
by a continuous symmetry-reduced trajectory, whereas in the next step a
\PoincSec\ replaces a \emph{continuous} time trajectory by a
\emph{discrete} sequence of points.

The main lesson of the visual tour undertaken above is that if a
dynamical problem has a continuous symmetry, the symmetry \emph{must} be
reduced before any detailed analysis of the flow's \statesp\ geometry can
take place. So far, this has only been achieved for transitionally
turbulent numerical pipe flows,\rf{ACHKW11} resulting in the discovery of
the first \rpo s embedded in turbulence. In the future, it should be the
first step in the analysis of any turbulent data, numerical or
experimental. Once symmetry reduction is achieved, all
solutions of a turbulent flow can be plotted together: all
symmetry-equivalent states are represented by a single point, families of
solutions are mapped to a single solution, \reqva\ become \eqva, \rpo s
become \po s, and most importantly, the analysis of the global dynamical
system in terms of invariant solutions and their stable/unstable
manifolds can now commence.

\begin{acknowledgments}
This article addresses the questions asked after the talk given at
Kyoto 2011 IUTAM Symposium on `50 Years of Chaos: Applied and Theoretical'.
We are indebted to
S.~Flynn,
S.~Froehlich,
J.~Greensite,
S.A.~Solla,
R.~Wilczak
and
A.P.~Willis
for inspiring discussions.
P.C.\ thanks G.~Robinson,~Jr.\ for support,
Max-Planck-Institut f\"ur Dynamik und Selbstorganisation,
G\"ottingen for hospitality,
and the Nieders\"achsischen Knackwurst and Bayerische Hefeweizen for
making it all possible.
P.C.\ was partly supported by NSF grant DMS-0807574
and
2009 Forschungspreis der Alexander von Humboldt-Stiftung.
D.B.\ thanks M.F. Schatz and was supported by NSF grant CBET-0853691.
\end{acknowledgments}

\end{document}